\documentclass[a4paper,11pt]{article}
\pdfoutput=1 

\usepackage{jheppub}
\usepackage{color} 
\usepackage{float}
\usepackage{ulem}
\usepackage{multirow}
\usepackage{cancel}
\allowdisplaybreaks

\usepackage[ddmmyyyy,hhmmss]{datetime}

\usepackage{amsmath,graphicx,amssymb,subfigure,bbm,mathtools,upgreek}
\usepackage[all]{xy}

\def\beq{\begin{equation}}
\def\eeq{\end{equation}}
\def\beqa{\begin{eqnarray}}
\def\eeqa{\end{eqnarray}}

\usepackage{todonotes}

\begin{document}

\title{Thermodynamics and DC conductivity of 2D anisotropic fluids from axion holography}

\author[a]{Alfonso Ballon-Bayona,}
\author[b,c,d]{Jonathan P. Shock}
\author[e,f]{and Dimitrios Zoakos}
\affiliation[a]{Instituto de F\'{i}sica, Universidade
Federal do Rio de Janeiro, \\
Caixa Postal 68528, RJ 21941-972, Brazil.}       
\affiliation[b]{Department of Mathematics and Applied Mathematics,
University of Cape Town, \\
Private Bag, Rondebosch 7700, South Africa.}
\affiliation[c]{The National Institute for Theoretical and Computational Sciences,
Private Bag X1, Matieland, South Africa.}
\affiliation[d]{Institut National de la Recherche Scientifique, Montreal, Canada}
\affiliation[e]{Department of Physics, University of Patras, 26504 Patras, Greece.}
\affiliation[f]{Department of Physics, 
National and Kapodistrian University of Athens, 15784 Athens, Greece.}

\emailAdd{aballonb@if.ufrj.br}
\emailAdd{jonathan.shock@uct.ac.za}
\emailAdd{dzoakos@upatras.gr}

\abstract{We investigate a strongly coupled finite-density anisotropic fluid  in $2+1$ dimensions dual to an  asymptotically AdS black brane that is a solution of Einstein-Maxwell-Axion theory in $3+1$ dimensions. Despite the anisotropy, the fluid thermodynamic properties align with those of a conformal fluid. Moreover, we show that the fluid is stable under the increase of the anisotropy parameter. Additionally, we analyse the DC conductivity of the anisotropic fluid, showing its compatibility with momentum dissipation due to translational symmetry breaking. In the limit of very large anisotropy we find that the DC conductivity vanishes as a consequence of dimensionality reduction. We also find that a  metal-insulator transition arises driven by the anisotropy.}

\maketitle

\flushbottom


\section{Introduction}
\label{Sec:Intro}

Gauge-Gravity duality provides a critical framework for exploring a diverse range of applications, from Quantum Chromodynamics (QCD) 
and condensed matter physics to quantum information theory (see e.g 
\cite{Ammon:2015wua,Hartnoll:2018xxg,Zaanen:2015oix}). 
For strongly coupled systems, where perturbative methods are inapplicable, it offers a powerful alternative framework for analysis. These situations are typically dominated by many-body collective dynamics, where elementary particle excitations are not well defined. The main idea behind the approach of applied holography is to find models which are as
close as possible to the experimental results, while maintaining
analytic or numerical control of the holographic (toy) model. A prominent application of holography in condensed matter physics is the study of holographic superconductors where the breaking of the $U(1)$ symmetry at low temperatures is mapped to the emergence of scalar hair in the higher-dimensional AdS black hole \cite{Hartnoll:2008vx,Hartnoll:2008kx,Horowitz:2010gk}. 

In condensed matter systems, translational and rotational symmetries are often broken, either explicitly or spontaneously or both\footnote{The presence of a source in the field theory side corresponds to explicit breaking, while a non-trivial vev for an operator (i.e. the condensate) characterizes the spontaneous breaking. When a purely spontaneous state combines with a source, we are in the pseudo-spontaneous regime. The distinction of the cases is due to the asymptotic behavior of the field that is responsible for the symmetry breaking.}  
(for a recent review see \cite{Baggioli:2022pyb}), and the systems are not Poincar\'e invariant. In the case of electron transport, the breaking of translational invariance, as described by the Drude model, is responsible for the finite conductivity that occurs in metals. 
Systems with translational invariance and finite density exhibit infinite conductivity as translational invariance implies that there is no momentum dissipation, and
as a result a constant electric field accelerates the charges indefinitely.  
Therefore, investigating the conductivity of real materials using holography necessitates understanding the breaking of spatial translations
and the construction of a holographic framework that incorporates momentum dissipation.  

In holography, translational symmetry breaking can be understood in terms of the Ward identity for the divergence of the holographic stress-energy tensor. In particular, for an Einstein-Maxwell-Scalar gravitational theory, the Ward identity becomes
\begin{equation} \label{Ward}
\nabla^{i} \langle T_{ij} \rangle = \partial_j \phi \langle O \rangle + F_{ji} \langle J^i \rangle
\end{equation}
where $i,j$ labels the boundary spacetime directions.  If the sources $\phi$ or $A_i$ become spatially dependent, the RHS of \eqref{Ward} will not vanish and will 
include contributions from the vacuum expectation value (vev) of either the scalar operator $\langle O \rangle$ or the 
current $\langle J_i \rangle$. The main message from \eqref{Ward} is that a possible route to momentum relaxation (and consequently to finite conductivity) is through spatially dependent sources. 

There are two classes of models with this characteristic: those where translational symmetry is broken while keeping the gravitational background homogeneous and those where the gravitational background becomes inhomogeneous.

In the inhomogeneous models, one introduces boundary conditions for a scalar field or the temporal component of a $U(1)$ gauge field that are spatially modulated, mimicking a lattice construction \cite{Horowitz:2012ky,Horowitz:2012gs,Horowitz:2013jaa,Donos:2014yya}, see also \cite{Nakamura:2009tf,Donos:2011bh}. Inhomogeneous models are usually technically challenging due to the need to solve coupled partial differential equations.

In the homogeneous models, one considers spatially dependent scalar sources that break translational symmetry while keeping the gravitational background homogeneous. This can be achieved by employing Einstein-Maxwell theory with a set of $d$ massless scalar fields, where $d$ is the number of spatial directions,  each field linear along one of the spatial directions, i.e.  $\chi_i = a \, x^i$. This is the isotropic analytical solution found in \cite{Andrade:2013gsa} that leads to the Drude-like DC conductivity
\begin{eqnarray}
\sigma_{DC} = r_h^{d-2} \Big [  1 + (d-1)^2 \frac{\mu^2}{a^2}   \Big ] \, ,  \label{DCconductivity}
\end{eqnarray}
where $r_h$ is the horizon radius and  $\mu$ the chemical potential. This approach, known as axion holography, can intriguingly be mapped to the massive gravity framework (see e.g. \cite{Vegh:2013sk,Davison:2013jba,Blake:2013bqa,Blake:2013owa}) where the graviton mass term breaks diffeomorphism invariance, see \cite{Alberte:2015isw}.  
The review \cite{Baggioli:2021xuv} comprehensively summarizes recent advances in this area, and contains many useful references. Note that in the limit $a \to 0$, where the scalar fields vanish, the DC conductivity is infinite and  translation invariance is restored. In the opposite limit $a \to \infty$ the DC conductivity reduces to $\sigma_{DC} = r_h^{d-2}$, which corresponds to incoherent charge transport \cite{Baggioli:2021xuv}.

This paper describes $2+1$-dimensional anisotropic fluids based on axion holography.
In contrast with \cite{Andrade:2013gsa}, we are interested in homogeneous models where the dissipation of momentum is related to the anisotropy of the gravitational background. To simplify the analysis we will turn on only one axion, i.e. $\chi = a \, x$, meaning that we are effectively anisotropising  the Hartnoll-Kovtun solution in \cite{Hartnoll:2007ai} 
(this is the limit of \cite{Andrade:2013gsa}  when both axions are set to zero). Notably, the anisotropy parameter $\alpha$ drives a non-trivial RG flow, culminating in the system's dimensional reduction. This RG flow will be manifest in many thermodynamic quantities and in the DC conductivity. In the conclusions we will comment on the anisotropy  in the two axion case and on the similarities/differences with respect to the analysis we present in this paper. For holographic descriptions of $3+1$ dimensional anisotropic fluids, see \cite{Mateos:2011tv,Jain:2014vka}.

In the axion holography framework, an alternative way to make the system anisotropic 
is to introduce mechanical deformation. If an arbitrarily large external strain is considered, the stress-strain 
relation becomes non-linear.  Common example of solids that exhibit such a behaviour are the rubbers, 
which more generally are referred as hyper-elastic materials. Holography has been used to model this kind of behaviour \cite{Baggioli:2020qdg,Pan:2021cux,Ji:2022ovs} and the deformation strain is introduced as an external field. 
The main difference between this approach and our model is that  
in these papers the holographic stress tensor is non-diagonal (due to the shear strain) whist in our case is diagonal. Another important difference is that with mechanical deformation translational symmetry is broken spontaneously  whereas in our case translational symmetry is broken explicitly.  Wherever appropriate we will compare some of our results with the findings of \cite{Baggioli:2020qdg,Pan:2021cux,Ji:2022ovs}.


The outline of the paper is as follows: In section \ref{Sec:anisotfluid} we describe the thermodynamic
properties of a conformal fluid in $D$ dimensions (specializing to the $2+1$ dimensional case) in the 
presence of finite charge density and anisotropy. 
In particular, we derive the equation of state and provide expressions for several thermodynamic quantities 
(including entropy, anisotropisation density, specific heat and susceptibility) in terms of the Helmholtz
free energy and its derivatives. Moreover, a thermodynamic stability analysis provides
a set of criteria that the susceptibility and  specific heat must obey, 
in order for the fluid to be stable. 

In section \ref{Sec:gravdual}  we describe the gravity dual of a $2+1$ 
dimensional charged anisotropic conformal fluid in terms of a gravity black brane solution that terminates on 
an $AdS_4$ boundary. Performing the standard holographic renormalisation procedure we obtain the 
Gibbs and Helmholtz free energies that are used to study the thermodynamic properties of the 
anisotropic fluid. From the components of the stress tensor, we compute the anisotropic enthalpy and pressures, and from these, we derive the speed of sound. From the perturbations of the background 
we extract the DC conductivity. The perturbative solution to the equations of motion for small values of the anisotropy is 
derived in appendix \ref{app-pert_solution} and it provides 
perturbative expressions for all the thermodynamic quantities as well as the conductivity. 

Section \ref{Sec:Numerics} is devoted to the numerical evaluation of the background 
and the conclusions that can be drawn from the study of the thermodynamic quantities and the DC conductivity, for 
generic values of the anisotropy and charge. 
Perturbative computations from appendix \ref{app-pert_solution} complement 
the numerical solutions in the limit of small values of the anisotropy, 
and appendix \ref{app:special_solutions} provides two 
analytic solutions that are conjectured to be the end points of the numerical solution in IR. 
The plot of the entropy density provides arguments in favor of the conjecture.  
From the plot of the speed of sound, an RG flow between a $D=2+1$ CFT at short distances and 
a $D=1+1$ CFT at large distances is unveiled, when the anisotropy is much larger than the charge. 
In appendix \ref{app-conductivity} we review the computation of the DC conductivity for holographic 
black holes that exhibit momentum dissipation.


\section{The anisotropic conformal fluid}
\label{Sec:anisotfluid}

This section discusses the thermodynamics of an anisotropic fluid exhibiting conformal symmetry. First we describe the equation of state of the fluid in $D$ spacetime  dimensions. For $D=3$, which is relevant to this work, we describe the canonical ensemble in which the charge density and anisotropy parameter are fixed. We end with a full description of the stability criteria for an anisotropic fluid in $2+1$ dimensions.


\subsection{Equation of state}

We analyse a conformal fluid at finite temperature $T$, with a finite charge density ${\cal Q}$ and anisotropy parameter $a$. The source of this anisotropy could come from different pressure gradients or from 
from unequal values for the spin interaction
along the two spatial directions. 
More generally, the anisotropy could arise as a result of some mechanism in which the $SO(2)$ rotational symmetry is broken. In our setup the anisotropy  arises from  considering a scalar source $\phi$ with a non-zero component of the gradient $\partial_i \phi$ along a particular spatial direction $x^i$.
In $D$ dimensions, the fluid's thermodynamics are governed by the following Gibbs free energy potential:
\begin{equation} \label{G-potential}
G = U - T S - (V {\cal Q}) \mu -  ( V  \Phi ) a = E - T S - (V {\cal Q}) \mu\, ,
\end{equation}
where $\Phi$ is the anisotropisation density and $\mu$ is the 
chemical potential. The anisotropisation density measures the response of the system to the anisotropy parameter.\footnote{This is analogous of the magnetisation density in the presence of a nonzero magnetic field} We have defined the anisotropic enthalpy as
\begin{equation} \label{aniso-enthalpy}
E = U -  ( V  \Phi ) a \, ,
\end{equation}
where $V$ is the volume, $U$ is the internal energy and $S$ is the entropy. 
From the first law of thermodynamics 
\begin{equation} \label{first-law-thermodynamics_v1}
d U = T d S - P dV + \mu \, d ( V {\cal Q} ) +  a \, d (V \Phi)\, ,
\end{equation}
and \eqref{G-potential}, we find that the thermodynamic evolution of the potential $G$ is given by
\begin{equation}  \label{Free-Energy-G_v1}
d G = - S dT - P d V - (V {\cal Q}) d \mu -  (V \Phi) d a \,.  
\end{equation}
For a conformal fluid we use the following dimensionless variables\footnote{\label{foot2}The pattern 
in the  notation we are going to follow is:  tilde when the dimensionless quantity is with respect to radius of the horizon 
(e.g. $ {\tilde a}=\frac{a}{r_h}$) and hat when it is with respect to the temperature (e.g. $\hat{a} = \frac{a}{T} $).} 

\begin{equation}  \label{Dimless}
{\hat \mu} = \frac{\mu}{T} \, ,  \quad 
{\hat a} = \frac{a}{T}  \, ,  \quad 
\hat {\cal Q} = \frac{ {\cal Q}}{T^{D-1}} \quad \& \quad 
{\hat \Phi}= \frac{\Phi}{T^{D-1}} \,.
\end{equation}
Conformal invariance and extensivity in $D$ dimensions constrains the potential to the following form
\begin{equation}  \label{ConfG}
G = V \,  T^{D} \, {\hat {\cal G}} ( {\hat \mu} , {\hat a})\, ,
\end{equation}
where $ {\hat {\cal G}} ( {\hat \mu} , {\hat a})$ is a potential that depends only on the dimensionless variables $\hat \mu$ and $\hat a$.

Now consider the scaling transformation $x' =  \lambda^{-1} x $, so that we obtain the rescaled quantities
\begin{equation}
T' = \lambda \, T \,  \quad 
V' = \lambda^{1-D} \, V \, , \quad  
\mu' = \lambda \, \mu  \, , \quad 
a' = \lambda \, a \quad \& \quad 
G' = \lambda \, G \,. 
\end{equation}
The conformal version of Euler's theorem takes the form
\begin{align}
G &= \Bigg [ \frac{d G'}{ d \lambda} \Bigg ]_{\lambda = 1} 
= \Bigg [ \frac{ \partial G'}{\partial T'} \frac{\partial T'}{ \partial \lambda} + \frac{ \partial G'}{\partial V'} \frac{\partial V'}{ \partial \lambda}  + 
\frac{ \partial G'}{\partial \mu'} \frac{\partial \mu'}{ \partial \lambda}  +  \frac{ \partial G'}{\partial a'} \frac{\partial a'}{ \partial \lambda} \Bigg ]_{\lambda=1} 
\nonumber \\
&= - S \, T + (D-1) P \, V - (V {\cal Q}) \,  \mu  -   (V \Phi) \, a \, . 
\end{align}
Using \eqref{ConfG}, the pressure can be written in terms of the potential as
\begin{equation}
P = - \frac{ \partial G} { \partial V} = - \frac{G}{V} \, .
\end{equation}
Then the conformal equation of state becomes
\begin{equation} \label{EQState}
D \, G = - S \, T  - (V {\cal Q}) \, \mu  -   (V \Phi) \, a 
\quad \xRightarrow[]{\text{$D$=3}} \quad
G = - \frac13 \, S \, T  - \frac13 \, ( V {\cal Q}) \, \mu - \frac13  \, (V \Phi) \, a \, . 
\end{equation}


\subsection{The canonical ensemble}

We will work in the canonical ensemble, where the charge density ${\cal Q}$ and the anisotropy parameter $a$ are fixed. The Helmholtz free energy is obtained through the Legendre transformation:
\begin{equation} \label{FGEq}
F =  G +  (V {\cal Q}) \, \mu = - \frac13 \, S \, T  + \frac23 \, ( V {\cal Q}) \, \mu - \frac13 \,  (V a) \, \Phi \,.
\end{equation}
We write the associated densities as
\begin{equation}
{\cal G} = \frac{G}{V} \, , \quad 
{\cal E} = \frac{E}{V} \,  , \quad  
{\cal U} = \frac{U}{V} \, ,  \quad 
 {\cal S} = \frac{S}{V} 
 \,\, \xRightarrow[]{\eqref{FGEq}} \,\,
 {\cal F} = \frac{F}{V}  = - \frac13 \, {\cal S} \, T  + \frac23 \, {\cal Q} \, \mu - \frac13 \, a \, \Phi\,.
\end{equation}
Thus, the evolution of the Helmholtz free energy density ${\cal F}$ takes the form
\begin{equation}
d {\cal F} = - {\cal S} \, d T  + \mu \, d {\cal Q} -  \Phi \, d  a \,. 
\end{equation}
We can write this free energy density as
\begin{equation}  \label{def_dimless_free_energy}
{\cal F} =  T^3  {\hat {\cal F}} (\hat {\cal Q}, {\hat a}) \quad {\rm with} \quad 
 {\hat {\cal F}} (\hat {\cal Q}, {\hat a}) = {\hat {\cal G}}(  {\hat \mu} ,{\hat a}) + \hat {\cal Q} \,  \hat \mu\, ,
\end{equation}
where we have used equations \eqref{ConfG} and \eqref{FGEq}  as well as the dimensionless variables given in 
\eqref{Dimless} for the case of $D=3$. In the next section we will use a dimensionless charge density $\hat Q$ related to $\hat {\cal Q}$ by $\hat {\cal Q} = 4 \, \sigma \, \hat Q$, where $\sigma$ is going to be defined through the gravity action.

From the dimensionless free energy ${\hat {\cal F}} (\hat {\cal Q}, {\hat a})$, we can obtain dimensionless ratios for the entropy density, chemical potential $\mu$, and anisotropisation density $\Phi$. 
We find
\begin{align} \label{dim-less-quantities}
\hat{\cal S} &= \frac{ {\cal S}}{T^2} 
= - \frac{1}{T^{2}}\,  \frac{ \partial {\cal F}}{\partial T} 
\Bigg{\vert}_{{\cal Q}, a}  = 
 - \, 3 \,  {\hat {\cal F}}    + \, 2 \,  {\hat {\cal Q}} \, 
 \frac{ \partial  {\hat {\cal F}}  }{ \partial {\hat {\cal Q}}} + 
{\hat a} \, \frac{ \partial  {\hat {\cal F}} }{ \partial {\hat a}}
\nonumber \\
\hat \mu &= \frac{\mu}{T} = \frac{1}{T} \, \frac{ \partial {\cal F}}{\partial {\cal Q}} \Bigg{\vert}_{T,a} 
= \frac{ \partial {\hat {\cal F}}  }{ \partial  {\hat {\cal Q}} } 
\quad \& \quad 
{\hat \Phi}= \frac{ \Phi}{T^2} = - \frac{1}{ T^{2}} \,  \frac{ \partial {\cal F}}{\partial a} \Bigg{\vert}_{T,{\cal Q}}
= - \, \frac { \partial  {\hat {\cal F}} }{ \partial {\hat a}}   \, . 
\end{align}
The free energy density is related to the anisotropic enthalpy density 
$ {\cal E}$ by
\begin{equation}
{\cal F} = {\cal E} - T {\cal S} \quad {\rm with} \quad {\cal E} = T^3 \, {\hat {\cal E}} ({\hat {\cal Q}}, {\hat a})\, ,
\end{equation}
where ${\hat {\cal E}} ({\hat {\cal Q}}, {\hat a})$ is the dimensionless anisotropic enthalpy density. 

The important quantities for analysing thermodynamic stability are the specific heat 
\begin{equation} \label{specific-heat_v1}
C_{{\cal Q},a} = T \frac{ \partial {\cal S}}{ \partial T } \Big{\vert}_{{\cal Q}, a} = 
T^2  \Bigg [ 2    - \, 2 \,  {\hat {\cal Q}} \, \frac{ \partial }{ \partial {\hat {\cal Q}}} - 
{\hat a} \, \frac{ \partial }{ \partial {\hat a}} \Bigg] {\hat {\cal S}} \, ,
\end{equation}
the charge and anisotropic susceptibilities \footnote{In this work we define the susceptibilities without taking the limit of the 
 corresponding variable going to zero. For example, in the charge susceptibility $\chi_{\mu}$ we do not take the limit $Q \to 0$.}
\begin{equation} \label{charge_anisotropic_susc}
\chi_{\mu} = \frac{\partial \mu}{\partial {\cal Q}} 
\Bigg{\vert}_{T,a} = \frac{1}{T} \, 
\frac{\partial^2  {\hat {\cal F}} }{\partial {\hat {\cal Q}}^2}    
\quad \& \quad 
\chi_{\Phi} = \frac{\partial \Phi}{\partial a} \Bigg{\vert}_{T, {\cal Q}} = - T \, \frac{ \partial^2  {\hat {\cal F}} }{ \partial {\hat a}^2} \, ,
\end{equation}
the pyro-charge and pyro-anisotropic coefficients
\begin{equation} \label{pyro_charge_anisotropic_v1}
\xi_{\mu} = \frac{ \partial \mu}{ \partial T} 
\Bigg{\vert}_{{\cal Q}, a} = 
\Bigg [ 1 - 2 \, \hat {\cal Q} \, \frac{\partial}{\partial \hat {\cal Q}} 
- \hat a \, \frac{\partial}{\partial \hat a} \Bigg ] \hat \mu 
\quad \& \quad
\xi_{\Phi} = \frac{ \partial \Phi}{ \partial T} 
\Bigg{\vert}_{{\cal Q}, a} = T \, 
\Bigg [ 2 - 2 \,  {\hat {\cal Q}} \, 
\frac{ \partial }{ \partial {\hat {\cal Q}}} - 
{\hat a} \, \frac{ \partial }{ \partial {\hat a}} \Bigg] {\hat \Phi} \, ,
\end{equation}
and the modified specific heat (at fixed anisotropisation density)
\begin{equation} \label{specific-heat_Phi}
C_{{\cal Q}, \Phi} = C_{ {\cal Q}, a} - 
T \, \frac{ \xi_{\Phi}^2}{ \chi_{\Phi}} \, .  
\end{equation}


\subsection{Stability criteria}
\label{SubSec:Stab_criteria}

To analyse thermodynamic stability, we examine variations in the internal energy density $\cal U$ under small perturbations of the extensive quantities ${\cal S}$, ${\cal Q}$, and $\Phi$.
Our analysis follows \cite{Landau:1980mil} and draws upon additional insights from \cite{Mateos:2011tv,Ballon-Bayona:2022uyy}. 
In thermodynamic equilibrium, the quantity
\begin{equation}
{\cal U} - T_0 \, {\cal S} - \Phi \, a_0 - \mu_0 \, {\cal Q}\, ,
\end{equation}
is at a minimum. Hence, for small deviations from equilibrium, the above-defined quantity must exhibit a positive change.
\begin{equation}
\delta {\cal U} - T_0 \, \delta{\cal S} - a_0 \, \delta \Phi - \mu_0 \, \delta {\cal Q} > 0 \, . 
\end{equation}
Expanding $ {\cal U} = {\cal U} ({\cal S}, {\cal Q}, \Phi)$ in a second-order Taylor series with respect to the extensive quantities, we find
\begin{align}
\delta {\cal U} &= \frac{\partial {\cal U}}{\partial {\cal S}} \delta {\cal S}  + \frac{\partial {\cal U}}{ \partial {\cal Q}} \delta {\cal Q} 
+ \frac{ \partial {\cal U}}{\partial \Phi} \delta \Phi + \frac12 \frac{ \partial^2 {\cal U}}{\partial {\cal S}^2} \delta {\cal S}^2 
+ \frac12 \frac{ \partial^2 {\cal U}}{ \partial {\cal Q}^2} \delta {\cal Q}^2 
 \nonumber \\
&
+ \frac12 \frac{\partial^2 {\cal U}}{ \partial \Phi^2} \delta \Phi^2
+ \frac{ \partial^2 {\cal U}}{\partial {\cal S} \partial {\cal Q}} \delta {\cal S} \delta {\cal Q} 
+ \frac{ \partial^2 {\cal U}}{\partial {\cal S} \partial \Phi} \delta {\cal S} \delta \Phi
+ \frac{ \partial^2 {\cal U}}{\partial {\cal Q} \partial \Phi} \delta {\cal Q} \delta \Phi \, .
\end{align}
Using the first law of thermodynamics, the first-order terms cancel, leaving the stability condition as
\begin{equation}
\frac{ \partial^2 {\cal U}}{\partial {\cal S}^2} \delta {\cal S}^2 
+  \frac{ \partial^2 {\cal U}}{ \partial {\cal Q}^2} \delta {\cal Q}^2 
+  \frac{\partial^2 {\cal U}}{ \partial \Phi^2} \delta \Phi^2 
+ 2 \frac{ \partial^2 {\cal U}}{\partial {\cal S} \partial {\cal Q}} \delta {\cal S} \delta {\cal Q} 
+ 2 \frac{ \partial^2 {\cal U}}{\partial {\cal S} \partial \Phi} \delta {\cal S} \delta \Phi
+ 2 \frac{ \partial^2 {\cal U}}{\partial {\cal Q} \partial \Phi} \delta {\cal Q} \delta \Phi > 0 \,. 
\end{equation}
This stability condition can be conveniently expressed in matrix form as
\begin{equation} \label{stabilitymatrix}
X^T \, A \, X  > 0 \quad {\rm with} \quad 
X = (\delta {\cal S} ,  \delta {\cal Q} , \delta \Phi) 
\quad \&  \quad 
A  = \begin{pmatrix}
  \frac{ \partial T}{ \partial {\cal S}}  & \frac{ \partial \mu}{ \partial {\cal S}} & \frac{ \partial a}{ \partial {\cal S}} \\
   \frac{ \partial T}{ \partial {\cal Q}}   &  \frac{ \partial \mu}{ \partial {\cal Q}} & \frac{ \partial a}{\partial {\cal Q}}  \\
\frac{ \partial T}{ \partial \Phi}  & \frac{ \partial \mu}{ \partial \Phi}  & \frac{ \partial a}{ \partial \Phi}  
\end{pmatrix}  \, . 
\end{equation}
Note that the thermodynamic variables for the energy density ${\cal U}$ are ${\cal S}$,  ${\cal Q}$ and $\Phi$ (all extensive quantities).
Condition \eqref{stabilitymatrix} is satisfied if all principal minors (determinants of leading submatrices) are positive definite (see, e.g., \cite{strang09}).
For the $1 \times 1$ determinant, we find
\begin{align} \label{1x1Det}
& A_{1,1} =   \frac{T}{C_{Q,\Phi}}  > 0 \, .
\end{align}
The $2 \times 2$ determinant evaluates to
\begin{align}  \label{2x2Det}
& A_{1,1} \, A_{2,2} - A_{1,2}^2 
=   \frac{T}{C_{Q,\Phi}} 
\Bigg [ \chi_{\mu} +  \frac{1}{{\chi_{\Phi}}} \left(\frac{\partial^2 {\cal F}}{\partial a \, \partial {\cal Q}} \right)_{T}^2\Bigg ] > 0 \, .
\end{align}
The $3 \times 3$ determinant is given by
\begin{equation} \label{3x3Det}
\det A = \frac{T \, \chi_{{\mu}}}{C_{{\cal Q}, \Phi} \, \chi_{\Phi}} 
 > 0 \, . 
\end{equation}
By combining \eqref{1x1Det}, \eqref{2x2Det}, and \eqref{3x3Det}, the stability criteria are expressed as
\begin{align}
 \chi_{\Phi} > 0  \, , \quad 
 \chi_{\mu} > 0 \quad  \& \quad 
C_{ {\cal Q}, \Phi}   > 0 \,. 
\label{stability_criteria}
\end{align}
Subsequent sections will verify whether these stability constraints hold using the explicit anisotropic gravity solution.


\section{Gravity dual of the anisotropic conformal fluid}
\label{Sec:gravdual}

This section introduces a charged anisotropic gravity black brane solution terminating at an $AdS_4$ boundary. The near-horizon (IR) and near-boundary (UV) asymptotics characterizing the black brane solution are discussed. Using holographic renormalisation, we evaluate the on-shell action and the holographic stress-energy tensor, deriving expressions for the free energy, anisotropic enthalpy, and pressures in terms of IR and UV coefficients. 

We will also use a perturbative analytical solution, valid for small anisotropy, to obtain results for the free energy and entropy densities. The details of this perturbative solution are given in appendix \ref{app-pert_solution}.
For generic values of the deformation parameters (charge and anisotropy), the equations of motion are solved numerically and the results are presented in section \ref{Sec:Numerics}. In that section we will also investigate the limit of large anisotropy and compare with our predictions based on special analytical solutions obtained in appendix \ref{app:special_solutions}.

\subsection{The asymptotically $AdS_4$ anisotropic brane}

We begin with the Einstein-Maxwell-Axion action in 4d
\begin{equation} \label{4d-action}
S = \sigma \int d^4 x \sqrt{-g} \Big [ R + \frac{6}{\ell^2} - \frac{c}{4} F_{\mu \nu}^2 - \frac{1}{2} (\partial \chi)^2  \Big ] \, ,
\end{equation}
where $F_{\mu \nu}=\partial_{\mu} A_{\nu} - \partial_{\nu} A_{\mu}$ is the gauge field strength, $\chi$ is the axion and 
$\sigma=1/(16 \pi G_4)$ with $G_4$ the 4d Newtonian gravitational constant. Unlike \cite{Andrade:2013gsa}, we restrict to a single axion, focusing on anisotropy-driven momentum dissipation along a single spatial direction. 
The axions in 4d are massless scalar fields in the bulk, that are dual 
to marginal operators in 3d with conformal dimension $\Delta =3$ \cite{Donos:2014cya,Blake:2015ina}.
The gauge coupling is set to $c=4 \ell^2$, and henceforth, we adopt units with $\ell=1$. The Einstein equations from \eqref{4d-action} take the form
\begin{equation} \label{einstein_v1} 
R_{\mu \nu} - \frac{1}{2} \, g_{\mu \nu}\, R - 3\, g_{\mu \nu} = 2 \, T^A_{\mu \nu} +  T^{\chi}_{\mu \nu}\, ,
\end{equation}
with 
\begin{equation}
T^A_{\mu \nu}  = F_{\mu \rho} \, F_{\nu}{}^{\rho} - \frac{1}{4} \, g_{\mu \nu} F_{pq} \, F^{pq} 
\quad \& \quad
 T^{\chi}_{\mu \nu} =\frac{1}{2}\Big[\partial_{\mu} \chi \, \partial_{\nu} \chi  - \frac{1}{2} \, g_{\mu \nu} (\partial \chi)^2 \Big]  \, . 
\end{equation}
Taking the trace of \eqref{einstein_v1} and substituting back, we derive the Einstein equations in the form
\begin{equation} \label{einstein_v2} 
R +12 = \frac{1}{2} \, (\partial \chi)^2 \quad \Rightarrow \quad 
R_{\mu \nu} +3 \, g_{\mu \nu} = 2 \, T^A_{\mu \nu} + \frac{1}{2} \, \partial_{\mu} \chi \, \partial_{\nu} \chi \, ,
\end{equation}
which are particularly useful for solving for the perturbations needed to compute the conductivity.

For the gauge field and the axion, the equations of motion are
\begin{equation} \label{Maxwell_axion_EOM}
D_{\mu} F^{\mu \nu} = \frac{1}{\sqrt{-g}} \partial_{\mu} \left(\sqrt{-g} \, F^{\mu \nu} \right) = 0 
\quad \& \quad 
D^2 \chi = \frac{1}{\sqrt{-g}} \partial_{\mu} \left(\sqrt{-g} \, g^{\mu \nu} \, \partial_{\nu} \chi \right) = 0 \, . 
\end{equation}
The ansatz for the metric, gauge field, and axion is given by
\begin{equation} \label{background_ansatz}
ds^2 =   - U(r) \, dt^2  + \frac{dr^2}{U(r)} + e^{2 \, V(r)} d x^2 + e^{2 \, W(r)} d y^2 
\quad {\rm with} \quad 
A_{\mu} = f(r) \, dt 
\quad \& \quad \chi = a \, x \, . 
\end{equation}
As usual, the radial coordinate ranges from the horizon radius ($r=r_h$, where $U(r_h)=0$) 
to the boundary ($r\rightarrow \infty$). The linear dependence of the axion in the $x$ coordinate results in translational symmetry breaking along that spatial direction.\footnote{Notice that this is a non-periodic deformation and is different from the periodic deformation of the Q-lattice approach \cite{Donos:2013eha}, that is due to relevant deformations of the CFT.}
Additionally, the $x$-dependent axion induces anisotropy between the $x$ and $y$ coordinates. This is seen in the presence of different functions in front of the $d x^2$ and $ d y^2$ terms of the metric ansatz in \eqref{background_ansatz}.
Plugging the ansatz of \eqref{background_ansatz} into the equations of motion \eqref{einstein_v2} and 
\eqref{Maxwell_axion_EOM} we obtain the following set of differential equations
\begin{eqnarray} \label{system-EOMs}
& 2 \, U' \left(V'-W'\right)+4 \, U \left(V''+V'^2\right)+  a^2 \, e^{-2 V} = 0 &
\nonumber \\[5pt]
& V''+V'^2+W''+W'^2 =0 &
\\[5pt]
&U''+U' \left(V'+W'\right)- 2 \, Q^2 \, e^{-2 \left(V+W\right)} -6 =0&
\nonumber \\[5pt]
&4 \, Q^2  \, e^{-2 \left(V+W\right)}+2\, U' \left(V'+W'\right)+4 \, U \, V' \, W'+ a^2 \,  e^{-2 \, V}-12=0&
\nonumber 
\nonumber \\[5pt]
&f' = Q \, e^{-\left(V+W\right)}&
\nonumber \, ,
\end{eqnarray}
where the prime denotes differentiation with respect to $r$, and  $Q$ is a constant related to the charge density, ${\cal Q}$, defined in the previous section by ${\cal Q} = 4 \sigma Q$. 
An analytical check shows that one of the first three second-order differential equations in $r$ is redundant, as it can be derived by combining the other two with the fourth first-order constraint equation.

For the perturbative solution analysed in Appendix \ref{app-pert_solution}, we introduce the following convenient linear combinations of $U$ and $W$
\begin{equation} \label{change-function}
V= \frac{1}{2} \left(V_+  + V_-\right) \quad \& \quad  
W=  \frac{1}{2} \left(V_+  - V_-\right) \, ,
\end{equation}
in order for the equations of motion to decouple after the expansion in small $a$. 
The equations in \eqref{system-EOMs} become
\begin{eqnarray} \label{system-EOMs-v2}
& 2 \, U' \, V_-'+ U \Big[2\left(V_+'' +V_-''\right) +V_+'^2 + V_-'^2 + 2 \, V_+' \, V_-' \Big]+  a^2 \, e^{- V_+ - V_-} = 0 &
\nonumber \\[5pt]
& 2\, V_+''+V_+'^2+V_-'^2 =0 &
\\[5pt]
&U''+U' \, V_+' -  2 \, Q^2 \, e^{-2 \, V_+} -6 =0&
\nonumber \\[5pt]
&4 \, Q^2  \, e^{-2 \, V_+}+2\, U' \, V_+' + U \left(V_+'^2 - V_-'^2 \right)+ a^2 \,  e^{- V_+ - V_-} -12=0&
\nonumber 
\nonumber \\[5pt]
&f' = Q \, e^{- V_+} \, . &
\nonumber 
\end{eqnarray}
Integrating the last equation in \eqref{system-EOMs-v2} and imposing the boundary conditions $f(r_h)=0$ and $f(\infty)= \mu$, 
we obtain the following expression for the function $f(r)$ and the chemical potential $\mu$
\begin{equation}
f(r) = \mu +  Q \int_{\infty}^r e^{-V_{+}(y)} dy
\quad {\rm with} \quad \frac{\mu}{Q} = \int_{r_h}^{\infty} e^{-V_{+}(y)} dy \, . 
\end{equation}
A useful relation is obtained by combining the equations in \eqref{system-EOMs-v2}
\begin{equation}
\partial_r \Bigg\{ e^{V_+} \Big[ U'  + U \left(V_-' - V_+' \right) \Big]\Bigg\} \, = \, 4\, Q^2 \, e^{-\, V_+}\, .
\end{equation}
This can be readily integrated using the last equation in \eqref{system-EOMs-v2}, 
to give the following constraint
\begin{equation} \label{constraint}
{\cal C} \, = \, e^{V_+} \Big[ U'  + U \left(V_-' - V_+' \right) \Big] - 4\, Q \, f \, . 
\end{equation}
\begin{flushleft}
{\bf Asymptotics}
\end{flushleft}
Near the horizon $r=r_h$, the metric functions in the ansatz can be expanded in a Taylor series as
\begin{align} \label{IR-expansion}
U(r) &= U_{h,1} (r-r_h) + \frac{ 8 \, Q^2 + a^2 \, w_{h,0}^2}{4 v_{h,0}^2 w_{h,0}^2} \, (r-r_h)^2 + {\cal O} (r- r_h)^3 
\nonumber \\
\frac{e^{V(r)} }{v_{h,0}} & = 1 -  \frac{ 2\, Q^2 +\left(a^2 - 6 \, v_{h,0}^2\right) w_{h,0}^2 }{2 \, U_{h,1} \, v_{h,0}^2 \, w_{h,0}^2} (r - r_h ) 
+ \frac{a^2 Q^2}{4\, U_{h,1}^2 \, v_{h,0}^4 \, w_{h,0}^2} \, (r-r_h)^2 + {\cal O} (r-r_h)^3 
\nonumber \\
\frac{e^{W(r)}}{w_{h,0}} &= 1 - \frac{2 \, Q^2 - 6 \, v_{h,0}^2 \, w_{h,0}^2 }{2 \, U_{h,1} \, v_{h,0}^2 \, w_{h,0}^2} (r-r_h)
- \frac{a^2 \, Q^2}{4 \, U_{h,1}^2 \, v_{h,0}^4 \, w_{h,0}^2} \,  (r-r_h)^2 + {\cal O} (r-r_h)^3
\nonumber \\
\frac{f(r) }{Q} &= \frac{r - r_h}{v_{h,0} \, w_{h,0}} + \frac{ 4\, Q^2  + \left(a^2 - 12 \, v_{h,0}^2\right) 
w_{h,0}^2}{4 \, U_{h,1} \, v_{h,0}^3 \, w_{h,0}^3} \, (r- r_h)^2 
+ {\cal O} (r- r_h)^3  \, . 
\end{align}
The independent coefficients are $U_{h,1}$, which is related to the temperature, and $v_{h,0}$ and $w_{h,0}$, which are linked to the entropy.
Later we will use the gauge freedom in the $x$ and $y$ coordinates to fix the value of $v_{h,0}$.  $U_{h,1}$ and $w_{h,0}$ will then be the only free parameters when integrating numerically from the horizon 
to the boundary. Near the boundary, the metric functions have the following asymptotic expansion
\begin{align}  \label{UV-expansion}
\frac{U(r)}{r^2} &= 1 + \frac{U_{\infty,1}}{r} + \frac{1}{4\, r^2} \left( U_{\infty,1}^2 - a^2 \right) + \frac{U_{\infty,3}}{r^{3}}
+\frac{24\,  Q^2 - a^4 - 12 \, U_{\infty,1} \, U_{\infty,3}}{24\, r^{4} } + {\cal O}(r^{-5}) 
\nonumber \\
\frac{e^{V(r)}}{r} &= 1 + \frac{U_{\infty,1}}{2\, r} + \frac{a^2}{8 \, r^2} + \frac{v_{\infty,3}}{r^{3}}
+ \frac{1}{r^4}\left[\frac{13}{384} \, a^4 - \frac{1}{32} \,  a^2 \, U_{\infty,1}^2 - U_{\infty,1} \, v_{\infty,3} \right] + {\cal O}(r^{-5}) 
\nonumber \\
\frac{e^{W(r)}}{r} &= 1 + \frac{U_{\infty,1}}{2\, r} - \frac{a^2}{8\, r^2} - \frac{v_{\infty,3}}{ r^{3}}
+ \frac{1}{r^4}\left[ - \frac{11}{384} \,  a^4 + \frac{1}{32} \, a^2 U_{\infty,1}^2 + U_{\infty,1} \, v_{\infty,3} \right] + {\cal O} (r^{-5})
\nonumber \\
\frac{f(r) - \mu}{Q} &=  - \frac{1}{r} + \frac{U_{\infty,1}}{2\, r^2} - \frac{U_{\infty,1}^2}{4\, r^{3}} + \frac{U_{\infty,1}^3}{8\, r^{4}} 
- \frac{a^4 + 30 \, U_{\infty,1}^4}{480\, r^{5}}+ {\cal O}(r^{-6}) \, .
\end{align}
The coefficient of the leading $r^2$ term in the $U$ expansion is $1$ and this was already fixed by the differential equations. 
Contrary to that, the coefficients of the leading $r$ terms in the $e^{V}$ and $e^{W}$ expansions have originally arbitrary and are fixed to unity imposing $AdS$ asymptotics. The independent UV coefficients are $U_{\infty,1}$, $U_{\infty,3}$, and $v_{\infty,3}$

By combining the asymptotic results for the constant of motion near the horizon  (from \eqref{IR-expansion})  
and near the boundary (from \eqref{UV-expansion}) we obtain from the constraint \eqref{constraint} the following condition
\begin{equation}  \label{condition}
3  \, U_{\infty,3} + 6  \, v_{\infty,3} + \frac38 \, a^2 \, U_{\infty,1} + 4\,  Q \, \mu  
=  - \, U_{h,1} \, v_{h,0} \, w_{h,0} \, ,
\end{equation}
that relates  boundary to horizon data.
\begin{flushleft}
{\bf The isotropic solution ($a=0$)}
\end{flushleft}
For the specific case $a=0$, the solution is analytical and takes the form
\begin{equation} \label{HK_solution}
f=\frac{Q}{r_h} \left[1 - \frac{r_h}{r} \right] \, , \quad 
V= W = \ln r \quad \& \quad 
\frac{U}{r^2} =1 +\frac{Q^2}{r^4} - \left[1+\frac{Q^2}{r_h^4}\right] \, \frac{r_h^3}{r^3}\, ,
\end{equation}
where the integration constant for $f$ has been fixed by the requirement that $f(r_h)=0$.
This corresponds to the Reissner-Nordström-AdS planar-symmetric black brane, as described by Hartnoll and Kovtun in \cite{Hartnoll:2007ai}\footnote{Such 
black holes have a long history \cite{Kottler:1918cxc,Romans:1991nq,Chamblin:1999hg,Stephani:2003tm}.}, which is the solution that we  anisotropise by the inclusion 
of momentum relaxation along the $x$ coordinate.
Expanding the above expressions close to the boundary and close to the horizon we obtain
the following values for the asymptotic coefficients
\begin{eqnarray}  \label{HK_parameters}
&& U_{\infty,1} = v_{\infty,3} = 0 \quad \& \quad \frac{U_{\infty,3}}{r_h^3} = -1 - \frac{Q^2}{r_h^4} 
\nonumber\\[5pt]
&& \frac{v_{h,0}}{r_h}= \frac{w_{h,0}}{r_h} =1  \, , \quad \frac{U_{h,1} }{r_h} = 3 - \frac{Q^2}{r_h^4} \quad \& \quad \mu=\frac{Q}{r_h} \, . 
\end{eqnarray}
\begin{flushleft}
{\bf Analytical solution for small anisotropy}
\end{flushleft}
The equations of motion can be solved perturbatively for small values of the anisotropy parameter, $a$, and in appendix \ref{app-pert_solution} we present the analytic computation. 
We expand around the analytic Hartnoll-Kovtun background of \cite{Hartnoll:2007ai}. Expanding this 
solution close to the boundary and close to the horizon, we obtain the following expressions for the asymptotic coefficients 
\begin{eqnarray}  \label{Pert-solution_parameters}
&&\frac{U_{\infty, 1}}{a^2 \, r_h}= -\, V_{-2}(r_h)\, , \quad 
\frac{U_{\infty, 3}}{r_h^3} = - \, 1 - \frac{Q^2}{r_h^4} + \frac{a^2}{4 \, r_h^2}
\Bigg[1+2\, r_h^2 \left(3-\frac{Q^2}{r_h^4} \right)V_{-2} (r_h)\Bigg]
\nonumber\\[5pt]
&& 
\frac{v_{\infty, 3}}{a^2\, r_h} =- \,  \frac{1}{12}  \, , \quad 
\frac{v_{h,0}}{r_h}=1 \, , \quad 
\frac{\mu \, r_h}{Q} \, =\, 1+ \frac{a^2}{2} V_{-2} (r_h)
\\[5pt]
&& 
\frac{w_{h,0}}{r_h} =1 -  a^2 \, V_{-2}(r_h) \quad \& \quad 
\frac{U_{h,1}}{r_h} = 3 - \frac{Q^2}{r_h^4} 
- \frac{a^2}{4 \, r_h^2}
\Bigg[1+6\, r_h^2 \left(1+\frac{Q^2}{r_h^4} \right)V_{-2} (r_h)\Bigg]\, ,
\nonumber
\end{eqnarray}
where $V_{-2}(r_h)$ is the function defined in \eqref{Vm2analytic} 
(or equivalently by the integral expression in \eqref{Vm2integral}), 
evaluated at $r=r_h$. Notice that it depends on the charge density constant
$Q$ and the horizon radius $r_h$.
It is important to note that the above expressions are valid for small anisotropy parameter $a$ but arbitrary charge density $Q$. This allows us to 
approach two distinct areas of the parametric space, namely small $a$ \& small $Q$ 
and small $a$ and large $Q$. We will explore those two limits for the 
computation of the different physical quantities later in this section. 

The perturbative solution agrees fully with the numerical solution to the equations of motion, as presented in the next section.
Moreover, the perturbative 
solution \eqref{Pert-solution_parameters} trivially satisfies the condition  \eqref{condition}
between horizon and boundary data.


\subsection{Holographic renormalisation and the stress-energy tensor}

To renormalize the action in \eqref{4d-action}, it is necessary to include a Gibbons-Hawking term and counterterms to cancel boundary divergences.

Using \eqref{einstein_v2}, The Euclidean version of the on-shell bulk action is given by
\begin{equation}
S_{\cal M} =  - \, \sigma \int d^4 x \sqrt{-g} \, \left [ R + 6 - F_{\mu \nu}^2 - \frac12 (\partial \chi)^2  \right] = 
\sigma \int d^4 x \sqrt{-g} \, \Big[ 6+  F_{\mu \nu}^2 \Big] \, .
\end{equation}
Using the ansatz for the metric and the gauge field, the action above becomes
\begin{equation}
S_{\cal M} =   
\sigma \, V_2 \, \beta  \int dr \, e^{V+W} \, \Big[ 6-  2 \,Q^2\,  e^{-2\left(V+W\right)} \, \Big] \, . 
\end{equation}
By combining the third and fifth equations in \eqref{system-EOMs}, the integrand can be expressed as a total derivative
\begin{equation}
\frac{S_{\cal M}}{\sigma \, V_2 \, \beta} =   
\int_{r_h}^{r_0} dr \, \partial_r \Big[ e^{V+W} \, U' - 4 \, Q \, f \, \Big] = 
\Big[ e^{V+W} \, U' \Big]_{r=r_h}^{r=r_0} - 4 \, Q \, f(r_0)\, ,
\end{equation}
since the horizon function $f$ obeys the condition $f(r_h)=0$. The boundary is defined at $r=r_0$ and we will take the limit $r_0 \to \infty$. The surface term is given by the Gibbons-Hawking term
\begin{equation}
S_{\partial \cal M} =  - 2 \, \sigma \, \int_{\partial \cal M} dx^3 \sqrt{\gamma} \, K = \sigma \, V_2 \, \beta \Bigg[e^{V+W} \, U'  -2 \Big(U \, e^{V+W}\Big)' 
\Bigg]_{r=r_0} \, . 
\end{equation}
Summing the two contributions and using the perturbative expansion close to the boundary from \eqref{UV-expansion} we arrive at the following expression
\begin{eqnarray}
\frac{S_{\cal M} + S_{\partial \cal M} }{ - \sigma \, V_2 \, \beta} & =& 4 \, r_0^3 + 6 \, U_{\infty, 1} \, r_0^2 +
\left(3 \, U_{\infty, 1}^2- a^2\right) r_0 
\\ 
&+& 
\Bigg[4 \, Q \, \mu - \frac{1}{2}\, a ^2  \, U_{\infty, 1}
\, + \, 4 \, U_{\infty, 3} \, + \, \frac{1}{2} \, U_{\infty, 1}^3 \Bigg] \, + \,  U_{h,1} \, v_{h,0} \, w_{h,0} + 
{\cal O}(r_0^{-1})\, .
\nonumber 
\end{eqnarray}
Here, $\mu$ represents the value of the function $f$ at the boundary, i.e., $f(r_0) = \mu$.
Two counterterms are added to cancel the divergences, specifically 
\begin{eqnarray} \label{counter_action_1}
S^1_{ct} =  \sigma \int d^4 x \sqrt{\gamma} &=& \sigma \, V_2 \, \beta 
\Bigg\{r_0^3+\frac{3}{2} \, U_{\infty, 1} \, r_0^2 + \frac{1}{4}\, r_0 
\left(3 \, U_{\infty, 1}^2-\frac{a^2}{2}\right) 
\nonumber \\
&+&\frac{1}{8} \left(- \frac{1}{2}\, a^2 \, U_{\infty,1} + 4\, U_{\infty,3} +U_{\infty,1}^3\right) \Bigg\}\, ,
\end{eqnarray}
and
\begin{equation}  \label{counter_action_2}
S^2_{ct} =  \sigma \int d^4 x \sqrt{\gamma} \, (\partial \chi)^2 = 
\sigma \, V_2 \, \beta \, a^2 \, \left(r_0+ \frac{1}{2}\, U_{\infty, 1} \right)\, .
\end{equation}
Consequently, the renormalised action is given by
\begin{equation} \label{Sren_v1}
S_{ren} = S_{\cal M} + S_{\partial \cal M} + 4 \, S^1_{ct} - \frac{1}{2} \,S^2_{ct} \, . 
\end{equation}
Substituting the corresponding expansions, we arrive at the following  
\begin{equation} \label{Sren_full}
\frac{S_{ren}}{ - \, \sigma \, V_2 \, \beta} =  2 \, U_{\infty, 3}+ 4 \, Q \, \mu +  \, U_{h,1} \, v_{h,0} \, w_{h,0}  \, . 
\end{equation}
It is important to note that the renormalised action of \eqref{Sren_full} will be 
identified in the next subsection with the thermodynamic potential ${\cal G}$ of the previous section, i.e. ${\cal G} = S_{ren}/ \beta \, V_2$. 

For small values of the anisotropy parameter we use the results of  \eqref{Pert-solution_parameters}, 
to arrive at the following expression for the renormalised action
\begin{equation} \label{Sren-perturbative}
\frac{S_{ren}}{ - \, \sigma \, V_2 \, \beta \, r_h^3} =  1 + {\tilde Q}^2+ \frac{ {\tilde a}^2}{4}
\Bigg[1-2 \left(3- {\tilde Q}^2 \right) V_{-2}({\tilde Q}) \Bigg]
\,\,\,\,  {\rm with} \,\,\,\,  {\tilde Q} = \frac{Q}{r_h^2} 
\,\,\,\,  \& \,\,\,\,  {\tilde a} = \frac{a}{r_h} 
\end{equation}
where $V_{-2}({\tilde Q})$ is the function we have defined in \eqref{Vm2analytic} evaluated at $r=r_h$.

The stress-energy tensor is given by the variation of the on-shell Einstein-Maxwell-axion action, written in 
Lorentzian signature, with respect to the source $\gamma_{\mu \nu}^{(0)}= r_0^{-2}\gamma_{\mu \nu}$
\begin{equation}
T^{\mu \nu}_{ren}  =  \frac{2}{\sqrt{-\gamma^{(0)}}} \frac{ \delta S_{ren}}{ \delta \gamma_{\mu \nu}^{(0)}} 
= \frac{2 \, r_0^5}{\sqrt{-\gamma}} \frac{ \delta S_{ren}}{ \delta \gamma_{\mu \nu}} 
= T^{\mu \nu}_{reg} + T^{\mu \nu}_{ct}
\quad {\rm with} \quad 
\gamma_{\mu \nu} = {\rm diag} \Big [ - U, e^{2V}, e^{2W}\Big ]_{r_0} 
\end{equation}
and the components $T^{\mu \nu}_{reg}$ and $T^{\mu \nu}_{ct}$ take the following form
\begin{equation} \label{Tmunu_Reg+ct}
\frac{T^{\mu \nu}_{reg}}{2\, \sigma \, r_0^5}
= K \, \gamma^{\mu \nu} - K^{\mu \nu}
\quad \& \quad 
\frac{T^{\mu \nu}_{ct}}{2 \, \sigma \, r_0^5} =  - \, 2\,  \gamma^{\mu \nu} 
- \frac{1}{2} \Big[ \partial^{\mu} \chi \,  \partial^{\nu} \chi  - \frac{1}{2} \, \gamma^{\mu \nu}  \, (\partial \chi)^2 \Big]
\end{equation}
where $K^{\mu \nu}$ is the extrinsic curvature tensor and 
$K=\gamma_{\mu \nu} K^{\mu \nu} $ is its trace. 
Using the background ansatz \eqref{background_ansatz} and the UV asymptotic behavior \eqref{UV-expansion}, the stress-energy tensor components are given by
\begin{eqnarray} 
&& {\cal E}= - \langle T^t{}_{t} \rangle = -2 \, \sigma \, U_{\infty,3} \, , \quad 
P_x = \langle T^1{}_{1} \rangle=  -\sigma \Bigg[U_{\infty,3} - 6\, v_{\infty,3} - \frac{3}{8} \, a^2 \, U_{\infty,1} \Bigg]
\nonumber \\
&& \quad \quad \& \quad P_y = \langle T^2{}_{2} \rangle=  -  \sigma 
\Bigg[U_{\infty,3}+ 6\,  v_{\infty,3} +  \frac{3}{8} \, a^2 \, U_{\infty,1}  \Bigg]
\label{StressTensor_components}
\end{eqnarray}
where $ {\cal E}$ is the anisotropic enthalpy density and $P_x$, $P_y$ are the pressures along the spatial
directions $x$ and $y$ respectively. Notice that the trace of the stress-energy tensor vanishes, i.e. 
\begin{equation}
\langle T^{\mu}_{\, \, \mu} \rangle \, = \,  - \,  {\cal E} + P_x + P_y \, =  \, 0
\end{equation}
a property that is present in the isotropic case (see e.g. \cite{Kim:2014bza}) and survives even when 
the background becomes anisotropic.


\subsection{Thermodynamics}
\label{thermodynamics}

The discussion of this subsection is complementary to the discussion we presented in section \ref{Sec:anisotfluid}, 
focused on the specific gravitational action we are considering. 
The fluid is characterized by the entropy density ${\cal S}$ (conjugate to the temperature $T$), 
charge density $\cal Q$ (conjugate to the chemical potential $\mu$) and 
anisotropisation density $\Phi$  (conjugate to the anisotropy parameter $a$), all of which are extensive quantities.

Notice that we follow the magnetic description in terms of the axion, in which the anisotropy $a$ plays the role of the external magnetic field
and the $\Phi$ the role of the magnetization \cite{Andrade:2013gsa,Jain:2014vka} and is in contrast to the electric description that is followed in \cite{Mateos:2011tv,Cheng:2014qia}. The reason behind this choice is that with the electric description the system turns out to be thermodynamically unstable, for any value of the anisotropy. Moreover, from dimensional analysis one finds that $a$ has dimension of mass which suggests the thermodynamic interpretation of the anisotropy as an intensive variable and $\Phi$ the  conjugate density.

The first law of thermodynamics governs changes in the internal energy density due to variations in the extensive quantities ${\cal S}$,  ${\cal Q}$ and $\Phi$ (see also \eqref{first-law-thermodynamics_v1}) 
\begin{equation} \label{first-law-thermodynamics}
d {\cal U} = T\, d{\cal S} + \mu \,  d{\cal Q} \, + a \, d\Phi
\end{equation}
and after the Legendre transformation ${\cal G} = {\cal U} - T \, {\cal S} - {\cal Q} \, \mu - a \, \Phi$, 
the  thermodynamic potential  is obtained. This satisfies the following relation
(see also \eqref{Free-Energy-G_v1}) 
\begin{equation}  \label{Free-Energy-G}
d {\cal G} = - {\cal S} \, dT -  {\cal Q} \, d\mu \, - \Phi \, da \, . 
\end{equation}
Notice that this potential is related to the on-shell gravity action, i.e. ${\cal G} = S_{ren}/ \beta \, V_2$ 
from equation \eqref{Sren_full}.  The thermodynamic potential in \eqref{Free-Energy-G} is a function of the temperature, the anisotropy parameter and the chemical potential. 

As described in section \ref{Sec:anisotfluid}, we will keep the anisotropy parameter $a$ and the charge density ${\cal Q}$ fixed. We are thus interested in the Helmholtz free energy density
\begin{eqnarray} \label{Free-Energy-F}
{\cal F} = {\cal G}  + {\cal Q} \, \mu = {\cal E} - T \, {\cal S} \,.     
\end{eqnarray}
\begin{flushleft}
{\bf Temperature, entropy and Helmholtz free energy}
\end{flushleft}
The gravitational computation yields the following expression for the temperature 
\begin{equation} \label{def-temperature}
T = \frac{U'(r_h)}{4 \, \pi} = \frac{U_{h,1}}{4 \, \pi} 
\end{equation}
while the entropy density is given by the Bekenstein-Hawking area formula
\begin{equation}  \label{def-entropy}
{\cal S} = 4 \, \pi \, \sigma \, e^{V(r_h)+W(r_h)} =  4 \, \pi \, \sigma\, v_{h,0} \, w_{h,0} \, . 
\end{equation}
For small values of the anisotropy we use the results in \eqref{Pert-solution_parameters} 
to obtain the following expressions for the temperature and the entropy density
\begin{equation} \label{T+S_small_tilde_a}
T = \frac{r_h}{4 \, \pi} \left\{3 - {\tilde Q}^2 - \frac{{\tilde a}^2}{4}
\Bigg[1+6  \left(1+ {\tilde Q}^2\right)V_{-2} ({\tilde Q})\Bigg]  \right\}
\quad \& \quad 
\frac{{\cal S} }{4 \, \pi \, \sigma \, r_h^2}= 1 -  {\tilde a}^2 \, V_{-2}({\tilde Q}) \, .
\end{equation}
The entropy density can also be derived using the on-shell action in \eqref{Sren-perturbative} and the variation of the thermodynamic potential $\cal G$ from \eqref{Free-Energy-G}. 
In fact, we have verified the equivalence between the Bekenstein-Hawking formula for the entropy density and the variation of the grand canonical $\cal G$ in \eqref{Free-Energy-G}  for any value of the anisotropy parameter using the numerical solution, which will be described in the next section.

Exploiting the presence of conformal symmetry and in the spirit of footnote \ref{foot2}, we introduce the following dimensionless ratios
\begin{equation} \label{dim-less-Q-S-a-Phi}
\hat{Q} = \frac{Q}{T^2}\, , \quad  {\hat {\cal S}} = \frac{\cal S}{T^2} \, , \quad 
{\hat a} = \frac{a}{T} \quad \& \quad
{\hat \Phi} = \frac{\Phi}{T^2} \, . 
\end{equation}
Expanding  the expressions in 
\eqref{dimless_ratios_all_Q} for small values of ${\tilde Q}$, we obtain an
analytic expression for the dimensionless ratio  $\hat {\cal S}$ in terms of $\hat Q$ and  $\hat a$ 
\begin{eqnarray} \label{SoverT2_largeTexpansion}
{\hat {\cal S}} &=&    \frac{64\, \sigma \, \pi^3}{9} \Bigg[ 1 + \frac{27}{128 \, \pi^4} {\hat Q}^2 
 - \frac{3645}{65536  \, \pi^8} {\hat Q}^4 + 
 \frac{59049}{2097152 \, \pi ^{12}} {\hat Q}^6  \cdots \Bigg] 
 \nonumber \\
 && + \,  \frac{2\, \sigma \, \pi}{3} \,  
{\hat a}^2 \, \Bigg[ 1-  \frac{81}{256 \, \pi^4} {\hat Q}^2 + 
\frac{15309}{65536  \, \pi^8} {\hat Q}^4 \cdots \Bigg] 
\end{eqnarray}
Expanding \eqref{dimless_free_energy_v1} for small values of $\tilde Q$, we obtain the following analytic expression for the dimensionless Helmholtz free energy density 
\begin{eqnarray}  \label{dimless_free_energy_v2}
{\hat {\cal F}} &= & - \, \frac{64\, \sigma  \, \pi ^3}{27}\Bigg[ 1 - \frac{81}{128 \, \pi^4} {\hat Q}^2 
+\frac{2187}{65536 \, \pi ^8}  {\hat Q}^4
-\frac{19683}{2097152  \, \pi ^{12}}  {\hat Q}^6 \Bigg] 
\nonumber \\
\qquad \qquad && 
 -\frac{2 \, \sigma \, \pi }{3} {\hat a}^2 \Bigg[1+ \frac{27}{256 \, \pi ^4} {\hat Q}^2  
 - \frac{2187}{65536 \,  \pi ^8} {\hat Q}^4 + \frac{295245}{16777216  \, \pi ^{12}}  {\hat Q}^6\Bigg] \, . 
\end{eqnarray}
In appendix \ref{app:zeroT} we present the results of the computations 
for  ${\hat {\cal S}}$, $\hat \Phi$ and ${\hat {\cal F}}$ in the limit 
of zero temperature. 
Alternatively, ${\hat {\cal S}}$ can be computed by substituting the perturbative free energy result from \eqref{dim-less-quantities} and using ${\hat {\cal Q}} = 4 , \sigma , {\hat Q}$.
\begin{flushleft}
{\bf Anisotropic enthalpy, pressures and speeds of sound}
\end{flushleft}
As described in \eqref{StressTensor_components}, we identify the $tt$ component of the energy momentum tensor with the anisotropic enthalpy density by ${\cal E}= - \langle T^t{}_{t} \rangle$\footnote{The notion 
of anisotropic enthalpy density in the presence of an anisotropy parameter is similar to the notion of the magnetic enthalpy density in the presence of the magnetic field defined in \cite{Ballon-Bayona:2022uyy}.}. Using the expression for the 
renormalised on-shell action in \eqref{Sren_full} together with the calculation 
of the enthalpy density from the \eqref{StressTensor_components}, the relation
${\cal Q} = 4 \, \sigma \, Q$ and the gravitational computations 
for the temperature and the entropy from \eqref{def-temperature} and 
\eqref{def-entropy}, we arrive at the following relation 
\begin{equation}
{\cal G} = {\cal E} - \mu \, {\cal Q} - T \, {\cal S} \, .    
\end{equation}
We can also combine the condition \eqref{condition} with 
the expressions for the $xx$ and $yy$ components of the energy momentum tensor in \eqref{StressTensor_components},  
to obtain the following relations for the pressures
\begin{equation} \label{pressures_action}
P_x = 2 \, {\cal E} - \mu \, {\cal Q} - T \, {\cal S} = {\cal E} + {\cal G}
\quad \& \quad 
P_y = - \, {\cal E} + \mu \, {\cal Q} + T \, {\cal S} = - \, {\cal G}
\end{equation}
with
\begin{equation} \label{pressure-sum-diff-all-a}
P_x+P_y =  {\cal E} \quad \& \quad 
P_x- P_y = 3 \, {\cal E} - 2\, \mu \, {\cal Q} -2\,  T \, {\cal S} \, . 
\end{equation}
Combining the equation of state in \eqref{EQState} with 
\eqref{G-potential} and \eqref{aniso-enthalpy}, we can write the anisotropic enthalpy 
in terms of extensive quantities and their conjugates as follows
\begin{equation} \label{enthalpy_v2}
{\cal E}  = \frac{2}{3} \, T\, {\cal S} + \frac{2}{3} \, \mu \, {\cal Q} 
- \frac{1}{3} \, a \, \Phi
\end{equation}
and substituting that expression in \eqref{pressure-sum-diff-all-a} 
it is possible to rewrite the difference between the pressures solely 
in terms of the anisotropy
\begin{equation} \label{pressures-diff-all-a_v2}
P_x -P_y = - \, a \, \Phi  \, . 
\end{equation}
Notice that these relations are exact, and hold for any value of the 
anisotropy and charge. The perturbative solution is in agreement with 
these expressions, and in figure \ref{Fig:PlotpxmpyovT3vsalph} we verify 
\eqref{pressures-diff-all-a_v2} using the full numerical solution. 

The next step in the thermodynamic analysis is the calculation of the speeds of sound in the anisotropic fluid. 
The anisotropy leads to two distinct directions for pressure wave propagation, each characterized by a different speed.
These calculations are performed with $a$ and $\cal Q$ held constant.

The expressions for the two speeds of sound read 
\begin{equation} \label{speedofsound__a_q_fixed}
c_{s,x}^2 \, = \, \left. \frac{\partial P_x}{\partial {\cal E}} \right|_{a, {\cal Q}} = 
\frac{\, \left. \frac{\partial P_x}{\partial T} \right|_{a,  {\cal Q}}}{\, \left. \frac{\partial {\cal E}}{\partial T} \right|_{a,  {\cal Q}} }  = 
\frac{{\cal S}+ {\cal Q}\, \xi_{\mu} - a \, \xi_{\Phi}}{C_{a, {\cal Q}} }
\quad \&  \quad 
c_{s,y}^2 \, = \, \left. \frac{\partial P_y}{\partial {\cal E}} \right|_{a, {\cal Q}} = 
1- c_{s,x}^2
\end{equation}
where we have used the relation between the two pressures from 
\eqref{pressure-sum-diff-all-a} to express the speed of sound along the $y$ direction 
in terms of the speed of sound along the $x$ direction. Notice that when translations are broken the speed of sound disappears at small wave-vector and only appears at $\kappa \gg \Gamma$, where $\Gamma$ is the momentum relaxation rate \cite{Baggioli:2021xuv}. It would be interesting to perform a quasinormal mode analysis in the anisotropic system, in order to compute the sound modes along the $x$ and $y$ directions. That calculation would show whether a relation of the form $\omega =c_s \, \kappa$ holds, at least along the 
$y$ direction.

Using \eqref{specific_heat_v2}, \eqref{pressure_x_deriv}, and \eqref{dimless_free_energy_v2}, we derive the following expression for the speed of sound along the $x$ direction, valid for small $\hat a$ and small $\hat Q$
\begin{equation} \label{speedofsound_smallq}
c_{s,x}^2 =  \frac{1}{2} - \left[\frac{3}{64 \, \pi^2} - \frac{81}{16384 \, \pi^5} \, {\hat Q}^2 + \frac{2187}{1048576  \, \pi^{10}} \, {\hat Q}^4 + \cdots \right] 
{\hat a}^2 \, .  
\end{equation}


\subsection{The DC conductivity}

In Appendix \ref{app-conductivity}, we reproduce the calculation of electrical conductivity in the presence of axions (i.e., momentum relaxation), originally performed in \cite{Andrade:2013gsa,Donos:2014cya} (see also \cite{Blake:2015ina}).
Since we introduce an axion only along the $x$ direction, the expression for the $xx$-component of the electrical conductivity is given by  \eqref{xx_yy-conductivities} 
\begin{equation} \label{sigma_xx_v1}
\sigma_{xx} =  e^{W - V} + \frac{4 \, Q^2}{a^2 \, e^{W + V} } \Bigg|_{r_h} = 
 \frac{w_{h,0}}{v_{h,0}}+ \frac{4\, Q^2}{w_{h,0} \, v_{h,0} \, a^2}\, . 
\end{equation}
While the second term in \eqref{sigma_xx_v1} can be expressed in terms of the entropy density and the dimensionless function ${\hat {\cal F}}$, the first term cannot. This is likely because the conductivity is derived from background perturbations rather than thermodynamics.

Using \eqref{Pert-solution_parameters} and expanding for small values of $\tilde a$ (for any  $\tilde Q$), 
we arrive at the following expression
\begin{equation} \label{sigma_xx_v2}
\sigma_{xx} = \frac{4 \, {\tilde Q}^2}{\tilde \alpha^2} +1 + 4\, {\tilde Q}^2 \, V_{-2}({\tilde Q}) 
- V_{-2}({\tilde Q}) \Big[1- 4\, {\tilde Q}^2 \, V_{-2}({\tilde Q})  \Big] {\tilde a}^2 \, . 
\end{equation}
Expanding for small values of $\tilde Q$, we obtain an analytic expression for the conductivity in terms of $\hat Q$ and $\hat a$
\begin{eqnarray} \label{sigma_xx_largeTexpansion}
\sigma_{xx}  &=&   \frac{9}{4\, \pi^2}  \frac{{\hat Q}^2}{{\hat a}^2} 
\Bigg[ 1 -  \frac{27}{128\, \pi^4} \,  {\hat Q}^2 \Bigg]+ 1 -  \frac{27}{128\, \pi^4} \,  {\hat Q}^2
+  \frac{5103}{32768\, \pi^8} \,  {\hat Q}^4 
\nonumber \\
&+& \frac{9\, c_{10}}{4\, \pi^2} \,  {\hat a}^2 
\Bigg[1 +\frac{9 \left(324 \, c_{10}^2-132 \, c_{10} + 36 \, c_{12}+13 \right)}{1024 \, \pi ^4 \, c_{10}}\, {\hat Q}^2 \Bigg] + \, \cdots
\end{eqnarray}
where the constants $c_{10}$ and $c_{12}$ (the first two terms in the expansion of the constant $c_1$ in \eqref{c1-pert}) are 
defined in \eqref{c1-components}.
Notice that 
in \eqref{sigma_xx_largeTexpansion}, since it is valid for small values of both 
$\hat Q$ and $\hat a$, the dominant terms are the first and the third, and the next more dominant terms are 
the second, the fourth and the fifth. 
Additionally, the leading term is proportional to ${\hat a}^{-2}$, which may appear related to the order of the perturbative ansatz in \eqref{change-function} and \eqref{perturbative_ansatz}. However that is 
not true, since the origin of the ${\hat a}^{-2}$  term is from the second term in \eqref{sigma_xx_v1} and even if the 
perturbative solution was until order ${\cal O}(a^{4})$, the dependence of that therm on $\hat a$ would be the same.


\section{Numerical results}
\label{Sec:Numerics}

This section presents the main results for the thermodynamics and DC conductivity of the conformal anisotropic fluid. First, we describe the numerical solution characterized by horizon parameters, which depend on the charge density and anisotropy parameter. Next, we present results for entropy, enthalpy, and Helmholtz free energy densities, comparing them to perturbative results from the analytical solution in Appendix \ref{app-pert_solution} for small anisotropy. We will also estimate the asymptotic behaviour of the entropy and the free energy in the limit of very large anisotropy and compare with the special analytical solution of appendix \ref{app:special_solutions}. We also analyse the chemical potential, anisotropization density, susceptibilities, specific heats, and speeds of sound. We interpret the latter in terms of an RG flow.  Lastly, we present our results for the DC conductivity and highlight an interesting novel behaviour for the conductivity at large anisotropy. 


\subsection{Numerical solution}

To numerically solve the independent Einstein-Maxwell-Axion equations in \eqref{system-EOMs}, we use dimensionless coordinates $\tilde r = r/r_h$ and $( \tilde t , \tilde x , \tilde y) = r_h (t,x,y)$. The corresponding dimensionless fields and  parameters are
\begin{equation}
\tilde U(\tilde r) = \frac{U(r)}{r_h^{2}} \, , \quad 
\left ( e^{\tilde V(\tilde r)} , e^{\tilde W (\tilde r)} \right )  = 
\left (\frac{e^{V(r)}}{r_h},  \frac{e^{W(r)}}{r_h} \right )  \, ,  \quad
\tilde f  = \frac{f}{r_h} \quad \& \quad
\tilde \mu  = \frac{\mu}{r_h} \, . 
\end{equation}
The definition for $\tilde Q$ and $\tilde a$ appears in \eqref{Sren-perturbative}.
The numerical solution is characterised by the parameters 
\begin{align}
\tilde v_{h,0}=r_h^{-1} v_{h,0} \,, \quad  
\tilde w_{h,0}=r_h^{-1} w_{h,0} \quad \& \quad 
\tilde U_{h,1}=r_h^{-1} U_{h,1} \, ,
\end{align}
where $v_{h,0}$, $w_{h,0}$ and $U_{h,1}$ are the horizon coefficients defined in \eqref{IR-expansion}. We use the scaling symmetry in $\tilde x$ to fix $\tilde v_{h,0}$ to $1$. The parameters $\tilde w_{h,0}$ and $\tilde U_{h,1}$ are determined by imposing AdS asymptotics near the boundary (large $ \tilde r$).

The dimensionless version of the temperature and entropy density given in \eqref{def-temperature} and \eqref{def-entropy} are 
\begin{align}
\tilde T = \frac{T}{r_h} \quad \& \quad 
\tilde  {\cal S} = \frac{S}{r_h^2} \quad
\xRightarrow[]{\text{where}} \quad 
\tilde T =  \frac{\tilde U_{h,1}}{4 \, \pi} \quad  \& \quad 
\tilde {\cal S} =  4 \, \pi \, \sigma\, \tilde v_{h,0} \tilde w_{h,0} \, . 
\end{align}
In order to map a dimensionless bulk quantity $\tilde X = r_h^{-n} X$ to a dimensionless physical quantity $\hat X$ we use the following identity
\begin{align}
\frac{\tilde X}{{\tilde T}^n} = \frac{X}{T^n} = \hat X \,. 
\end{align}

\begin{figure}[htb]
\centering
\includegraphics[width=6cm]{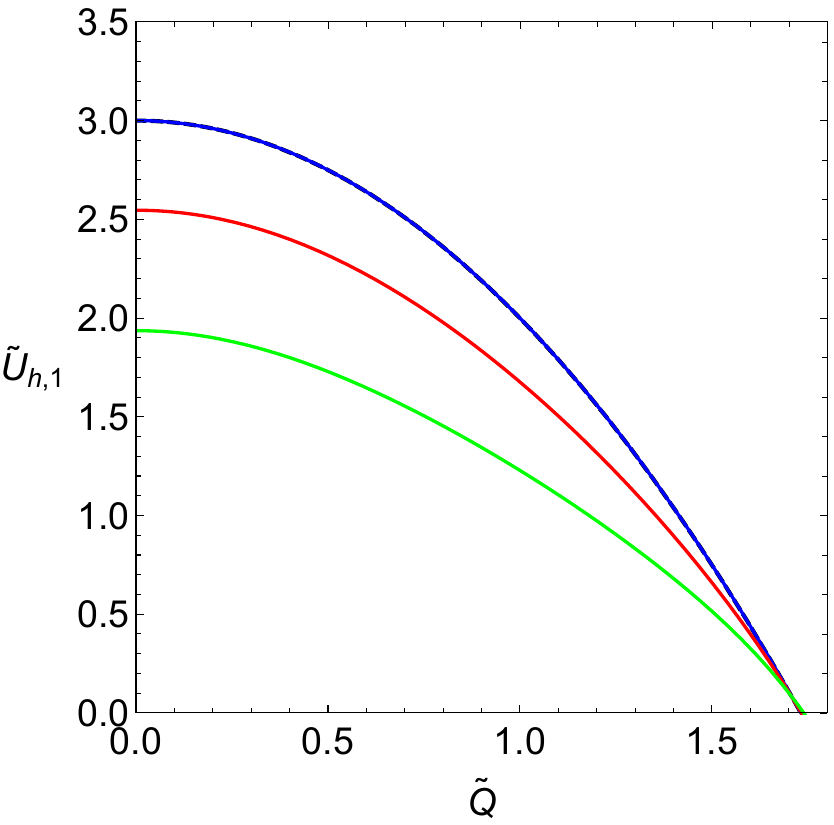}
\hspace{1cm}
\includegraphics[width=6cm]{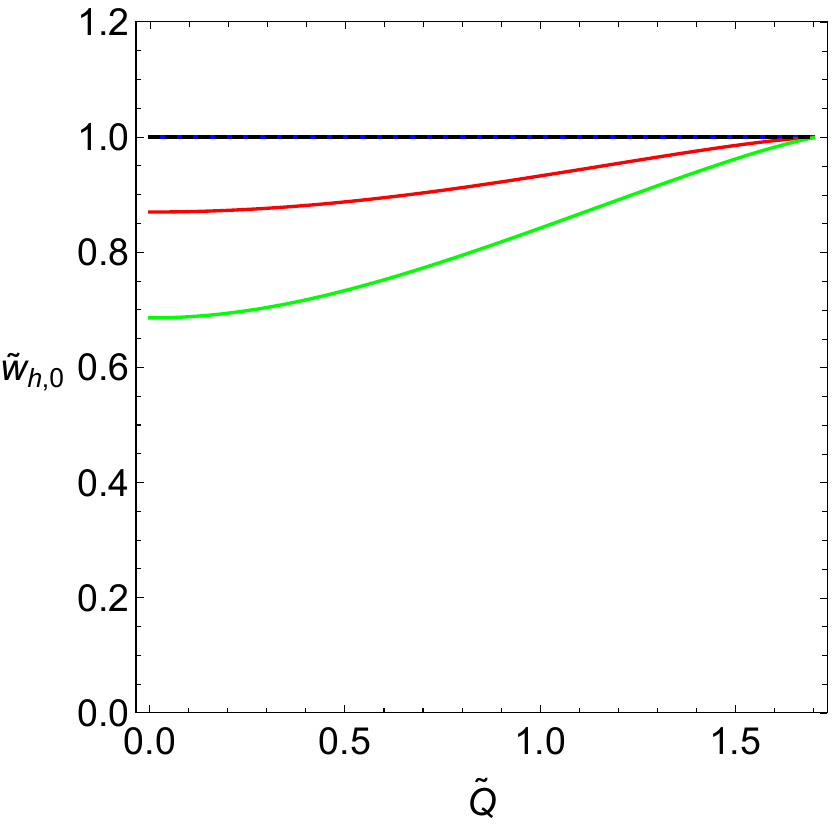}
\caption{The  horizon coefficients $\tilde U_{h,1}$ (Left) and $\tilde w_{h,0}$ (Right) as a function of the bulk parameter $\tilde Q$. The blue, red and green solid lines correspond to $\hat a =0$, $\hat a=5$ and $\hat a=10$ respectively where $\hat a = a/T$ is the dimensionless anisotropy parameter. The black dashed line is the analytical result for the isotropic case $\hat a=0$. Note that the maximum value of $\tilde Q$ is $\sqrt{3}$.}  
\label{Fig:PlotUh1vsTBQ}
\end{figure}

Our numerical results for $\tilde w_{h,0}$ and $\tilde U_{h,1}$ are shown in figure \ref{Fig:PlotUh1vsTBQ}. The curves represent the variation of $\tilde U_{h,1}$ and $\tilde w_{h,0}$  with $\tilde Q$ for fixed values of the dimensionless anisotropy parameter $\hat a$. The curve that corresponds to the value $\hat a =0$ in the 
right panel of figure \ref{Fig:PlotUh1vsTBQ} is a straight line parallel to the horizontal axis. For $\hat a=0$ the value of $\tilde w_{h,0}=1$ for any values 
of $\hat Q$. This is the Hartnoll-Kovtun solution that we summarise in 
equation \eqref{HK_solution}.

\begin{figure}[htb]
\centering
\includegraphics[width=6cm]{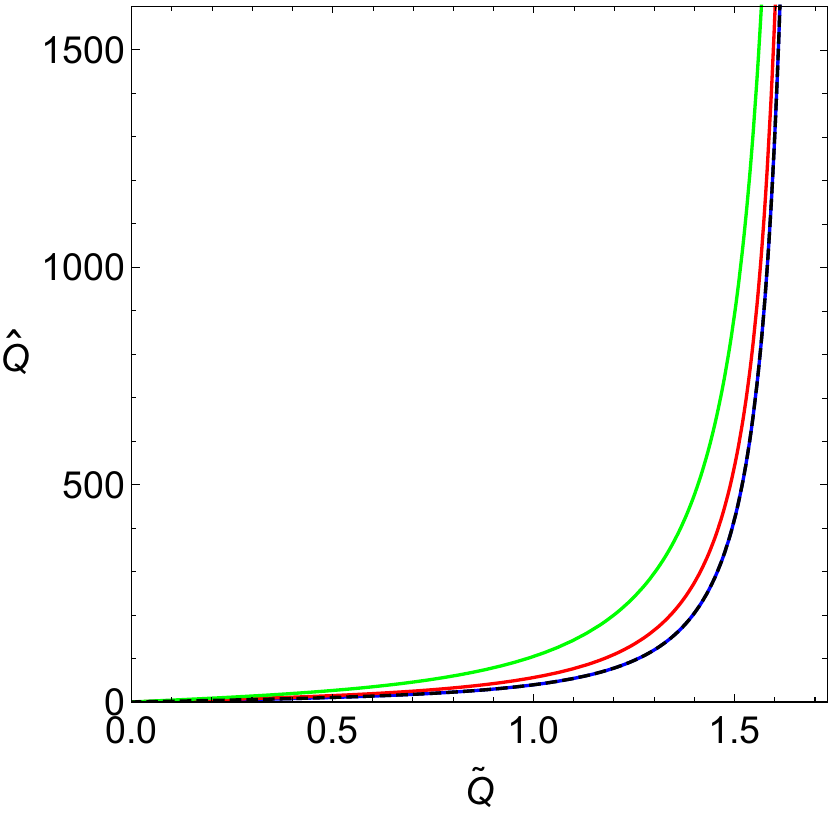}
\hspace{1cm}
\includegraphics[width=5.7cm]{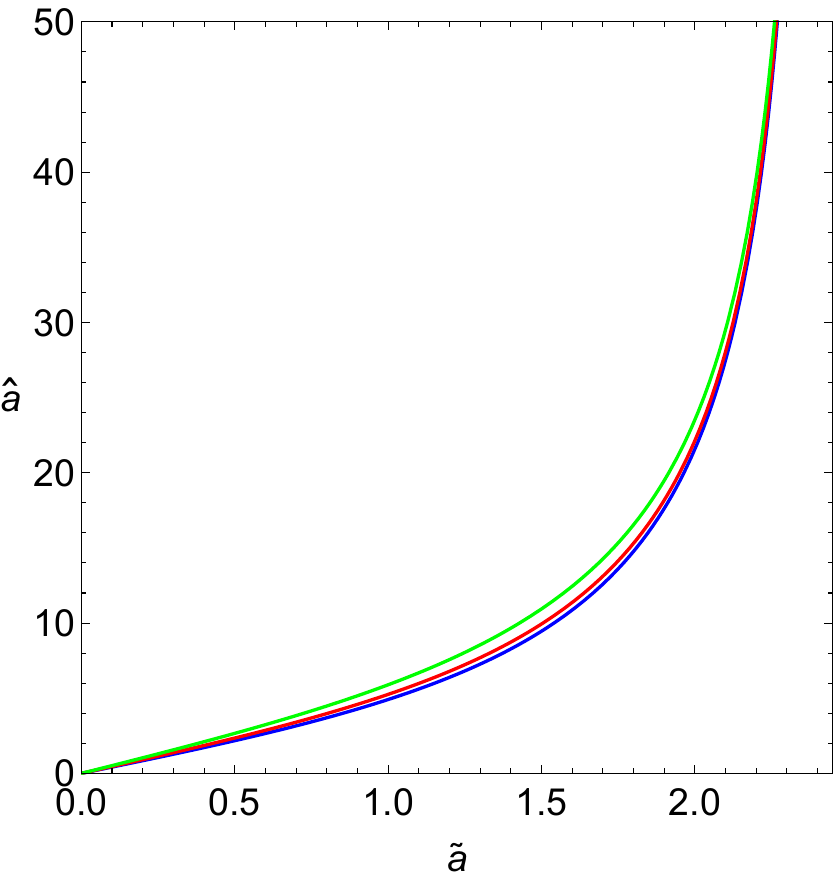}
\caption{{\bf Left:} The dimensionless charge density $\hat Q=Q/T^2$ as a function of the bulk parameter   $\tilde Q$.  The blue, red and green solid lines correspond to $\hat a =0$, $\hat a =5$ and $\hat a =10$ respectively where $\hat a = a/T$ is the dimensionless anisotropy parameter. The black dashed line represents the analytical solution for $\hat a =0$. The maximum value of $\tilde Q$ is $\sqrt{3}$ which corresponds to the limit $\hat Q \to \infty$. {\bf Right:} The dimensionless anisotropy parameter $\hat a$ as a function of the bulk parameter   $\tilde a$.  The blue, red and green solid lines correspond to $\hat Q =0$, $\hat Q =10$ and $\hat Q =20$ respectively. The maximum value of $\tilde a$ is $\sqrt{6}$ which corresponds to the limit $\hat a \to \infty$.}  
\label{Fig:PlotTQvsq}
\end{figure}

The bulk parameters $\tilde Q$ and $\tilde a$ vary monotonically with the dimensionless physical parameters $\hat Q$ and $\hat a$, respectively, as illustrated in Figure \ref{Fig:PlotTQvsq}. Note that the limits $\hat Q \to \infty$ and $\hat a \to \infty$ correspond to the limits $\tilde Q \to \sqrt{3}$ and $\tilde a \to \sqrt{6}$ respectively.
These limiting values are determined by the IR boundary conditions. In order to find numerical solutions that interpolate smoothly 
between the near horizon asymptotics and the $AdS_4$ boundary, the first 
order derivative of the functions $W$ and $V$ at the horizon should be positive. Looking in \eqref{IR-expansion} for the boundary behaviour of the function $W$ and after setting $\tilde{v}_{h,0}=1$, we end up with the condition $\tilde{Q} \le \sqrt{3} \, \tilde{w}_{h,0}$. Since we know (see also the right panel of figure \ref{Fig:PlotUh1vsTBQ}) that $\tilde{w}_{h,0} \le 1$ we conclude that  $\tilde{Q} \le \sqrt{3}$. In the same fashion from the boundary behaviour of the function $V$ we have that 
${\tilde a}^2 \le  6 - \tilde{Q} ^2 \, \tilde{w}_{h,0}^{-2}$. This expression attains its maximum value for $\tilde Q = 0$, leading to ${\tilde a} \le \sqrt{6}$.

\subsection{Entropy density}
\label{SubSec:Num_entropy}

\begin{figure}[htb]
\centering
\includegraphics[width=6cm]{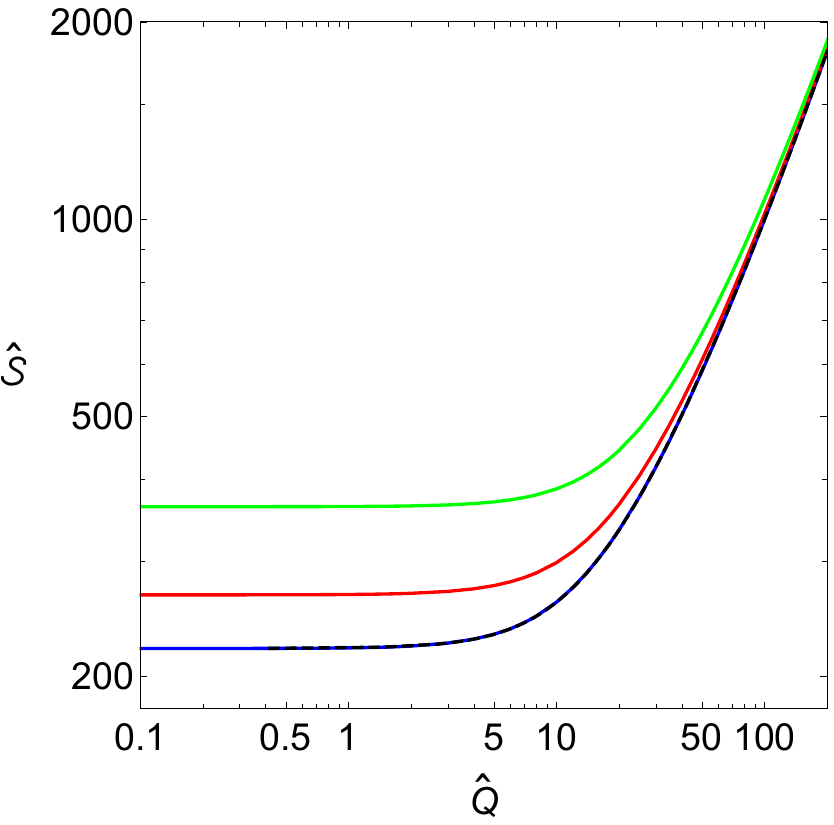}
\hspace{1cm}
\includegraphics[width=6cm]{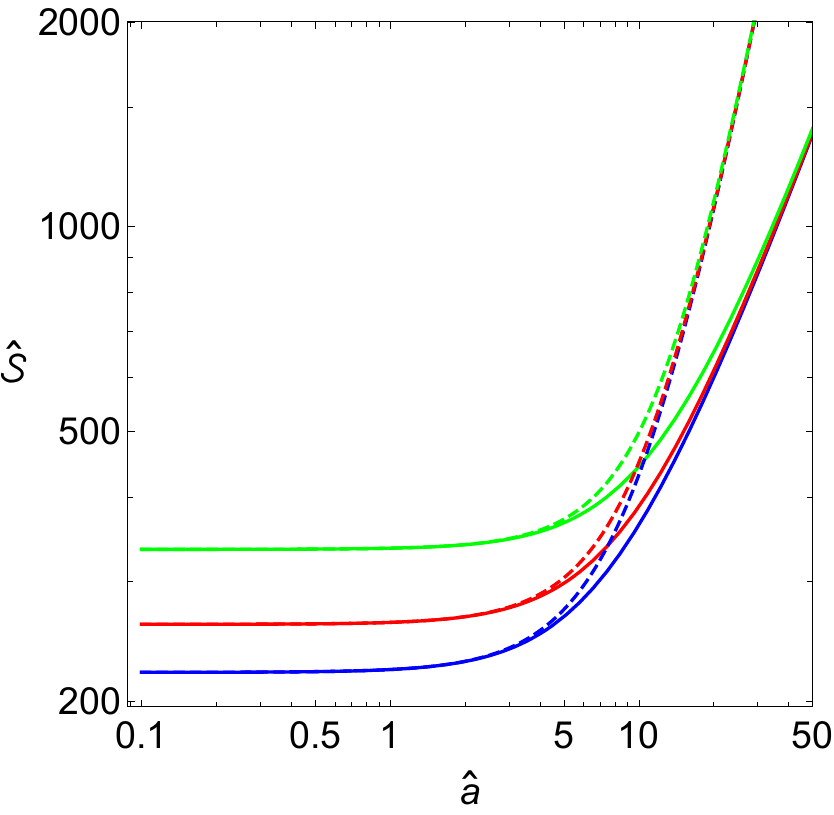}
\caption{{\bf Left Panel:} The dimensionless entropy density  $\hat {\cal S}= {\cal S}/T^2$ as a function of the dimensionless charge density $\hat Q=Q/T^2$. The blue, red and green solid lines  correspond to $\hat a=0$, $\hat a=5$ and $\hat a=10$ respectively  where $\hat a = a/T$ is the anisotropy parameter. The black dashed line on the left panel represents the analytical solution for $\hat a=0$. {\bf Right Panel:} The dimensionless entropy density  $\hat {\cal S}= {\cal S}/T^2$ as a function of the dimensionless anisotropy parameter $\hat a = a/T$. The solid lines represent the numerical result whilst the dashed lines represent the perturbative solution. The blue, red and green solid lines  correspond to $\hat Q=0$, $\hat Q=10$ and $\hat Q=20$ respectively. We have set the gravitational constant $\sigma$ to one. }  
\label{Fig:Plotsvsq}
\end{figure}

Figure \ref{Fig:Plotsvsq} shows the entropy density $\hat {\cal S}$ as a function of $\hat Q$ (left panel) and $\hat a$ (right panel).The gravitational constant $\sigma$ is set to one in these plots. 
The dashed lines in the right panel have been drawn using the following 
perturbative expression for the entropy density\footnote{This is the first 
of the expansions in \eqref{dimless_ratios_all_Q}.} 
\begin{equation} \label{dimless_ratios_all_Q_entropy}
{\hat {\cal S}} (\hat a \ll 1) = \frac{64\, \sigma \, \pi^3}{(3-{\tilde Q}^2)^2} 
\Bigg\{1+ \frac{1}{2}\, \frac{{\tilde a}^2}{3-{\tilde Q}^2}
\Big[1+ 8\, {\tilde Q}^2 \, V_{-2} ({\tilde Q})\Big]\Bigg\} 
\end{equation}
that comes after combining the two equations in \eqref{T+S_small_tilde_a} 
and expanding until second order in $\tilde a$. 
For small values of $\hat a$ and $\hat Q$ the perturbative solution 
takes the form in \eqref{SoverT2_largeTexpansion}.

Both plots demonstrate that when one deformation significantly exceeds the other, it dominates the entropy density behavior. In the left panel, as $\hat Q$ exceeds 50, the three solid lines for different values of $\hat a$ converge. A similar phenomenon is observed in the right panel when $\hat a$ exceeds 30. The entropy density exhibits similar scaling in these two regions. 

In appendix \ref{app:special_solutions}, we present two 
analytic solutions. The first one is for zero anisotropy (i.e. the background in 
\eqref{special_1}) and the second for zero charge density (i.e. the background in 
\eqref{special_2}). 
The corresponding entropy densities, expressed in dimensionless terms, are calculated in \eqref{special_1_entropy} and \eqref{special_2_entropy}. In both cases, the entropy density scales linearly with $\hat Q$ or $\hat a$, but with different slopes.\footnote{Notice that in the 
isotropic case of the two equal axions \cite{Andrade:2013gsa}, 
the dimensionless entropy 
density scales quadratically  with $\hat a$ in the limit of $\hat a \gg \hat Q$
(see \cite{Kim:2014bza}).} 
We argue that these two backgrounds are limits of 
the full numerical solution when either $\hat Q$ is much larger than 
$\hat a$, or vice versa. In particular, we conjecture that for very large anisotropy, the entropy behaves as 
\begin{equation}  \label{entropy_largealpha}
\hat {\cal S} (\hat a \gg 1) =  s_{\infty,1} \, {\hat a}  \quad , \quad s_{\infty,1} = \frac{16\, \sigma \, \pi^2}{3\, \sqrt{6}} \, . 
\end{equation}
Comparing the relations in \eqref{special_1_entropy} and \eqref{special_2_entropy} with numerical results for the entropy density reveals good agreement. In the plots we have set the gravitational constant $\sigma$ to one. The agreement could be enhanced by determining the subleading term, though this is challenging, as discussed in Appendix \ref{app:special_solutions}.

As analysed in Appendix \ref{app:special_solutions}, when $\hat Q \gg \hat a$, the numerical solution interpolates between the metric in \eqref{special_1} near the horizon and the $AdS_4$ metric near the boundary.
When an Einstein/Maxwell theory (in this limit we effectively set the axion to zero) 
is placed at a finite chemical potential, an 
RG flow is induced that leads to the emergence of a semi-locally critical 
$AdS_2 \times R^2$ in the near horizon region (for more details see the 
discussion and analysis of section 4.2 of \cite{Hartnoll:2016apf} and also \cite{Hartnoll:2018xxg}). 

Similarly, when $\hat a \gg \hat Q$, the numerical solution transitions from the metric in \eqref{special_2} to the $AdS_4$ metric near the boundary.
Motivated by \cite{DHoker:2009mmn}, this reflects a different RG flow from a $D=2+1$ CFT at short distances to a $D=1+1$ CFT at long distances, from the perspective of the boundary field theory.
In other words, when the anisotropy dominates over the charge the
flow is from $AdS_4$ to $AdS_3$. 
This observation aligns with findings in other gravity backgrounds with anisotropy, e.g., \cite{Gursoy:2020kjd}. The flow from \eqref{special_2} 
to $AdS_4$ is characterized by an effective reduction in the spatial dimension 
of the $AdS$ space, which will become evident through the calculation of the 
speeds of sound along the $x$ and $y$ directions. We postpone this 
discussion for subsection \ref{SubSec:Num_SpecHeat_SpeedSound}.

In a different framework that translational invariance is broken spontaneously, through 
the introduction of mechanical deformation, the entropy density presents a scaling similar to \eqref{entropy_largealpha}. More specifically, in \cite{Pan:2021cux} 
the authors study the effects of non-linear elasticity on the thermodynamics and the entropy for large
values of the strain (which is the anisotropy parameter in that set-up) presents a universal scaling law 
with the following form: $s \sim \epsilon^{\zeta}$, where  $\epsilon$ is the shear strain. Exponent 
$\zeta$ is a parameter that is related to the bulk potential\footnote{More details about the 
form of the potential and the 
explicit expression of the parameter $\zeta$ can be found in equations (S36) and (S42) 
of \cite{Pan:2021cux}.} and for a specific choice of that potential it becomes  $\zeta=1$. 
and the entropy scaling matches precisely with \eqref{entropy_largealpha}. 
This universality in the scaling of the entropy is unexpected, if one takes into account that in the case of 
\cite{Pan:2021cux} translational symmetry breaks spontaneously, while in the system we examine in this 
paper breaks explicitly. However, this result arises from the emergence of a special scaling symmetry 
close to the horizon (in the limit of strong anisotropy) that characterizes both physical systems. 
In our case for large anisotropy, the system flows to an $AdS_3$, as we discussed in the Appendix \ref{app:special_solutions}.  In the case examined
in \cite{Pan:2021cux}\footnote{The analysis of the special form of the metric for large anisotropy/strain is
discussed in \cite{Baggioli:2020qdg},  around equation (3.29).}, the metric for large strain in the limit that the parameter $\zeta$ becomes one and after a convenient change of variables becomes again $AdS_3$. 
In summary, the entropy density has the same scaling in these two very different set-ups, since the geometries in the limit of large anisotropy (or strain) preserve the same symmetries.

\subsection{Anisotropic enthalpy  and Helmholtz free energy densities}

The anisotropic enthalpy density ${\cal E}$ is the $tt$ component of the stress energy tensor and its holographic dictionary was obtained in \eqref{StressTensor_components}. This paper focuses on the canonical ensemble, where the Helmholtz free energy density, ${\cal F} = {\cal E} - T {\cal S}$, is the primary thermodynamic quantity. All other thermodynamic quantities can be derived from the Helmholtz free energy density, as detailed in section \ref{Sec:anisotfluid}.
The conductivity, however, cannot be expressed solely in terms of ${\cal F}$ (or equivalently ${\hat {\cal F}})$, as it depends on metric and gauge field perturbations.

\begin{figure}[htb]
\centering
\includegraphics[width=6cm]{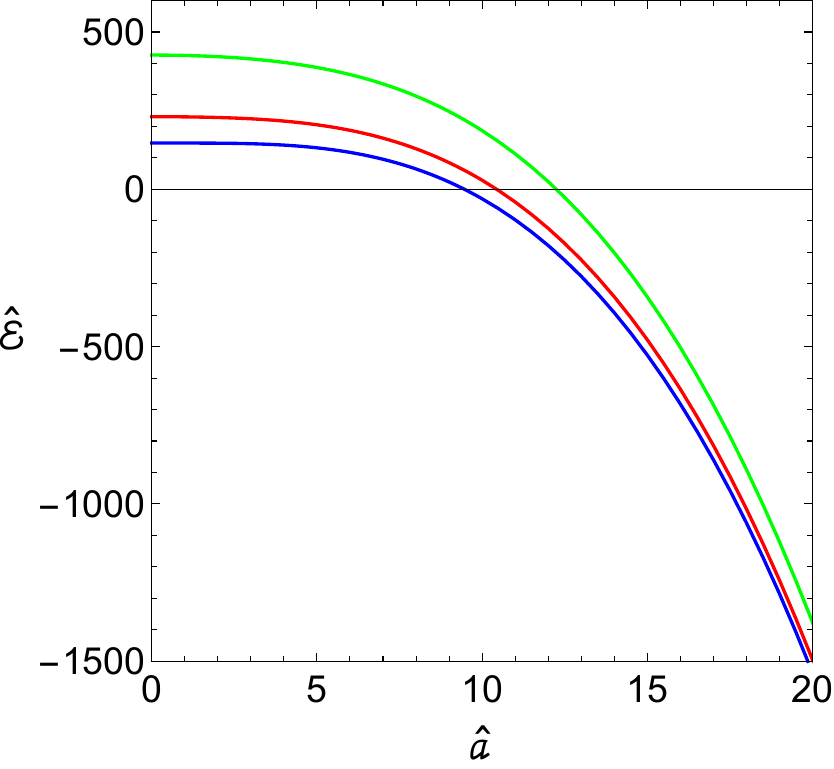}
\hspace{1cm}
\includegraphics[width=6cm]{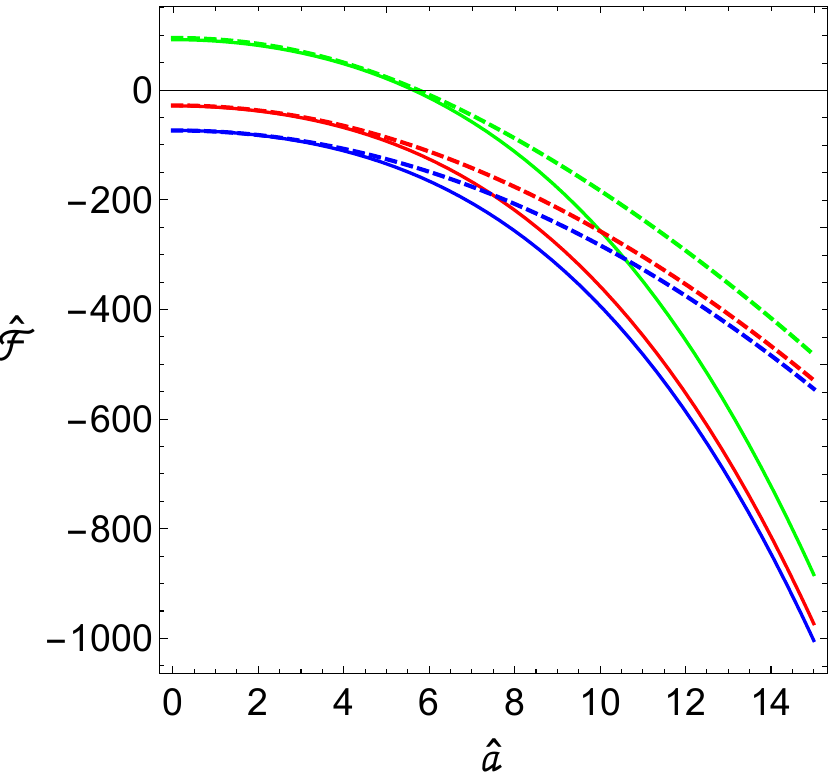}
\caption{{\bf Left Panel:} The dimensionless anisotropic enthalpy density $\hat {\cal E} = {\cal E}/T^3$ as a function of  the dimensionless anisotropy parameter $\hat a = a/T$. {\bf Right Panel:} The dimensionless Helmholtz free energy density $\hat {\cal F} = {\cal F}/T^3$ as a function of  the dimensionless anisotropy parameter $\hat a = a/T$.  In both panels the blue, red and green solid lines correspond to $\hat Q = Q/T^2$ equal to $0$, $10$ and $20$ respectively. The dashed lines in the right panel represent the perturbative result \eqref{Fsmallanisotropy}. We have set the gravitational constant $\sigma$ to one.}  
\label{Fig:Plotfvsalpha}
\end{figure}

Figure \ref{Fig:Plotfvsalpha} presents the dimensionless anisotropic enthalpy density ${\hat {\cal E}}$ (left panel) and Helmholtz free energy density ${\hat {\cal F}}$ (right panel). The gravitational constant $\sigma$ is set to one in the plots. Both quantities decrease monotonically with the anisotropy parameter $\hat a$, and the ${\hat {\cal F}}$ plot includes a comparison with perturbative calculations. The dashed lines are based on the following perturbative expression\footnote{This corresponds to equation \eqref{dimless_free_energy_v1} in appendix \ref{app-pert_solution}.}
\begin{equation}  \label{Fsmallanisotropy}
{\hat {\cal F}} (\hat a \ll 1) =  192 \, \sigma \, \pi^3 \, \frac{{\tilde Q}^2 -\frac{1}{3}}{(3-{\tilde Q}^2)^3} 
\Bigg\{1+ \frac{5}{6}\, {\tilde a}^2  \frac{{\tilde Q}^2-\frac{3}{5}}{(3-{\tilde Q}^2)\big({\tilde Q}^2-\frac{1}{3}\big)}
\Big[1+ \frac{24}{5}\, \frac{{\tilde Q}^2({\tilde Q}^2+1)}{ {\tilde Q}^2-\frac{3}{5} } \, V_{-2} ({\tilde Q})\Big]\Bigg\}
\end{equation}
where the function $V_{-2} ({\tilde Q})$ is defined in \eqref{Vm2integral} and \eqref{Vm2analytic}. For small $\hat a$ and $\hat Q$, the perturbative solution simplifies to the power series in \eqref{dimless_free_energy_v2}.
As with the entropy density summarized in figure \ref{Fig:Plotsvsq}, the perturbative solution accurately describes the system for small $\hat a$. 

For large $\hat a$, a polynomial fit to the numerical solution indicates that the leading behavior is
\begin{eqnarray}
\hat {\cal F} (\hat a \gg 1) = f_{\infty,3} \, \hat a^3 \quad , \quad  f_{\infty,3} \approx - 0.24 \,. 
\end{eqnarray}
The leading term does not contribute to the entropy density, as shown in eq. \eqref{dim-less-quantities}. A subleading term of the form $f_{\infty,1} \hat a$, where $f_{\infty,1} = - s_{\infty,1}/2$, contributes the leading term for entropy in \eqref{entropy_largealpha} and aligns well with numerical results.


\subsection{Chemical potential, anisotropisation density and susceptibilities}

\begin{figure}[htb]
\centering
\includegraphics[width=6cm]{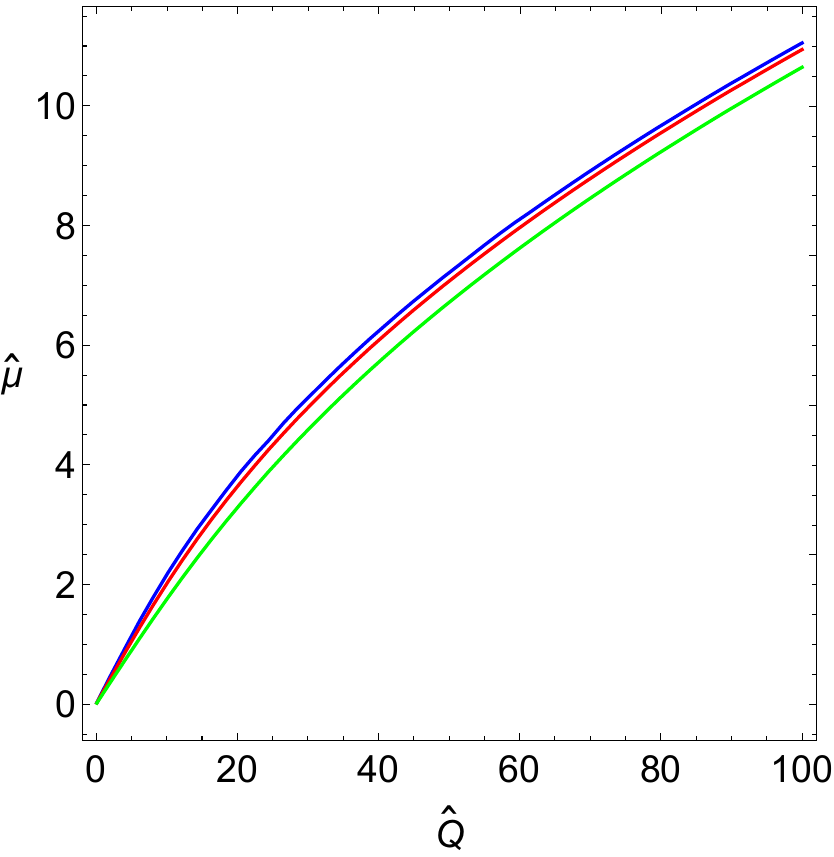}
\hspace{1cm}
\includegraphics[width=6cm]{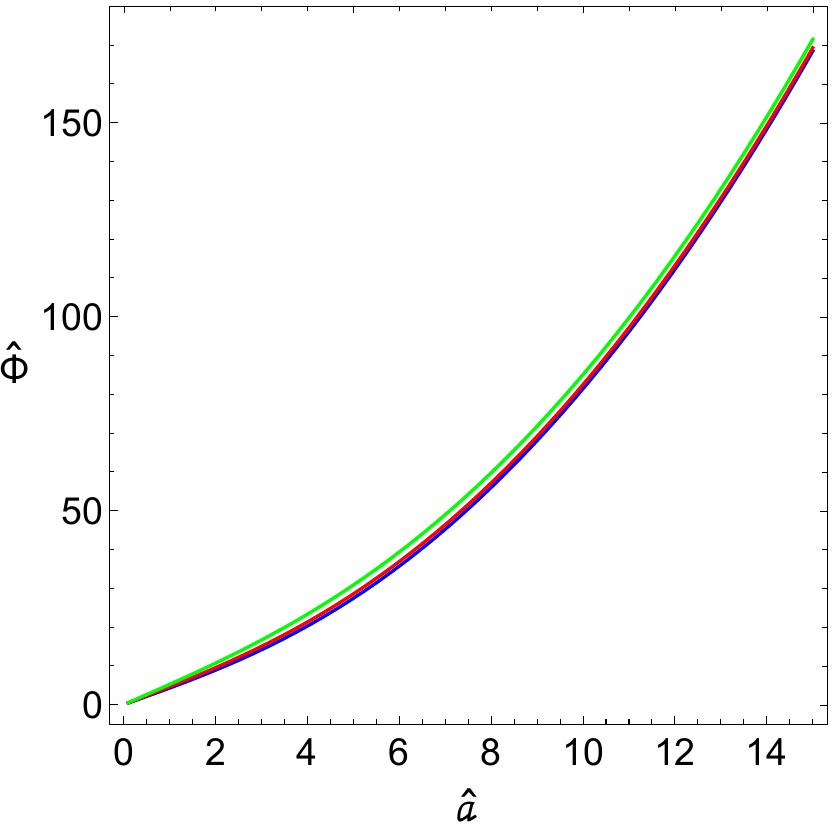}
\caption{{\bf Left Panel:} The dimensionless chemical potential $\hat \mu = \mu/T$ as a function of  the dimensionless charge density $\hat Q = Q/T^2$. The blue, red and green solid lines correspond to $\hat a = a/T$ equal to $0$, $5$ and $10$ respectively. {\bf Right Panel:} The dimensionless anisotropisation density $\hat \Phi = \Phi/T^2$ as a function of  the dimensionless anisotropy parameter $\hat a = a/T$. The blue, red and green solid lines correspond to $\hat Q = Q/T^2$ equal to $0$, $10$ and $20$ respectively. }  
\label{Fig:Plotmuvsqvphivsalpha}
\end{figure}

This work considers an ensemble where the charge density $Q$ and anisotropy parameter $a$ are fixed. The corresponding conjugate thermodynamic variables 
are the chemical potential $\mu$ and the anisotropisation density $\Phi$. 
In the left panel of figure \ref{Fig:Plotmuvsqvphivsalpha}, we plot the dimensionless 
chemical potential $\hat \mu$ as a function of $\hat Q$ for fixed values of $\hat a$.
We observe that the dependence of $\hat \mu$ on the different values of $\hat a$ is very mild and as you increase $\hat a$ the value of $\hat \mu$, for the same value of 
$\hat Q$, decreases. In the right panel of figure \ref{Fig:Plotmuvsqvphivsalpha}, 
we present the dimensionless anisotropisation density $\hat \Phi$ as a function 
$\hat a$ for fixed values of $\hat Q$. Similarly, $\hat \Phi$ depends mildly on $\hat Q$, and increases with $\hat Q$ for the same $\hat a$, as shown in the right panel.

\begin{figure}[htb]
\centering
\includegraphics[width=6cm]{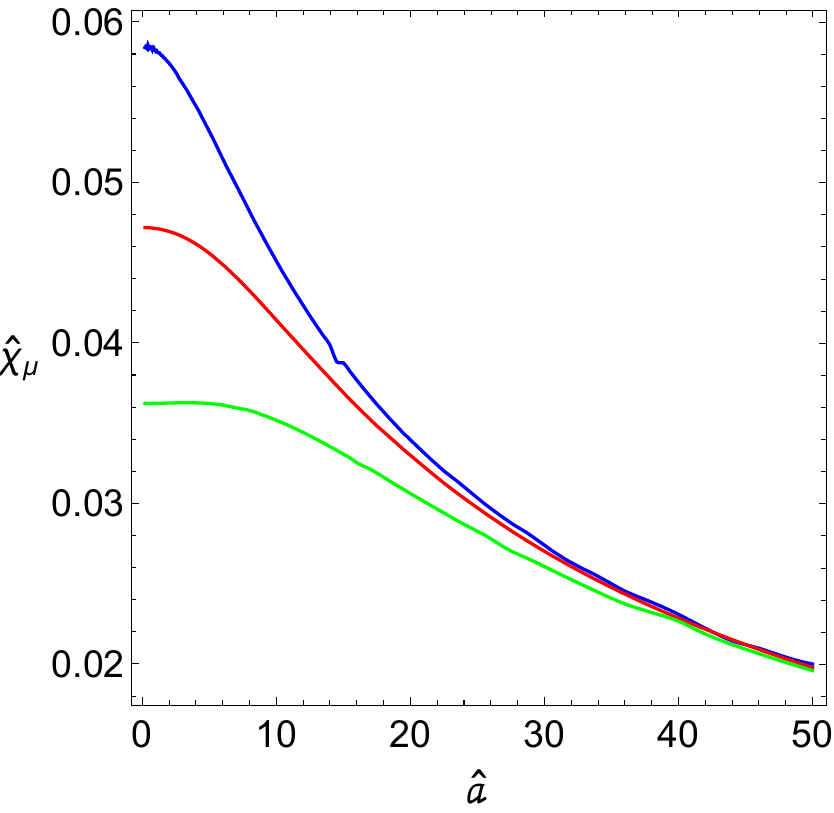}
\hspace{1cm}
\includegraphics[width=5.7cm]{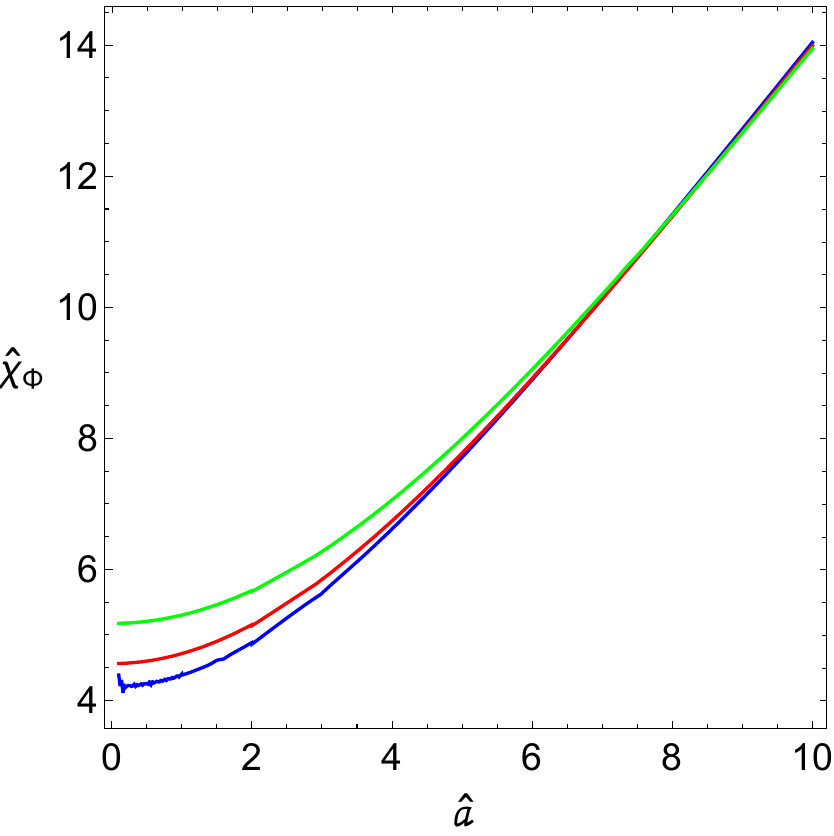}
\caption{ {\bf Left Panel:} The dimensionless charge susceptibility $\hat \chi_{\mu} = T \chi_{\mu}$ as a function of  the dimensionless anisotropy parameter $\hat a = a/T$.  {\bf Right Panel:} The dimensionless anisotropic susceptibility $\hat \chi_{\Phi} = \chi_{\Phi}/T$  as a function of  the dimensionless anisotropy parameter $\hat a = a/T$.  In both panels the blue, red and green solid lines correspond to $\hat Q = Q/T^2$ equal to $0$, $10$ and $20$ respectively. }  
\label{Fig:PlotchiPhiTchimuovTvsalpha}
\end{figure}

Both $\hat \mu (\hat Q)$ and $\hat \Phi (\hat a)$ increase monotonically, implying that their derivatives—related to charge and anisotropic susceptibilities (see equation \eqref{charge_anisotropic_susc})—are positive. In figure  
\ref{Fig:PlotchiPhiTchimuovTvsalpha} we plot the dimensionless charge and anisotropic susceptibilities. More precisely, in the 
left panel of figure \ref{Fig:PlotchiPhiTchimuovTvsalpha} we plot the 
dimensionless charge susceptibility $\hat \chi_{\mu}$ as a function of $\hat a$ and in the right panel 
the dimensionless anisotropic susceptibility $\hat \chi_{\Phi}$, also as a function of $\hat a$. 
The charge susceptibility decreases with $\hat a$, whereas the anisotropic susceptibility increases with $\hat a$, but both remain positive for all $\hat a$ and $\hat Q$. The calculation for small values of $\hat Q$ and 
$\hat a$ can be done using the perturbative solution and the results can be 
found in appendix \ref{app:SubSec:Stability}. More specifically, in 
\eqref{stability_condition_2_smallaQ} we present the calculation for the 
dimensionless charge susceptibility and in \eqref{stability_condition_3_smallaQ} the 
result for the anisotropic susceptibility. The perturbative results, combined with numerical computations, confirm that the first and second thermodynamic stability conditions in \eqref{stability_criteria} hold for all $\hat Q$ and $\hat a$. 

\subsection{Specific heats, speeds of sound and pressure anisotropy}
\label{SubSec:Num_SpecHeat_SpeedSound}

\begin{figure}[htb]
\centering
\includegraphics[width=6cm]{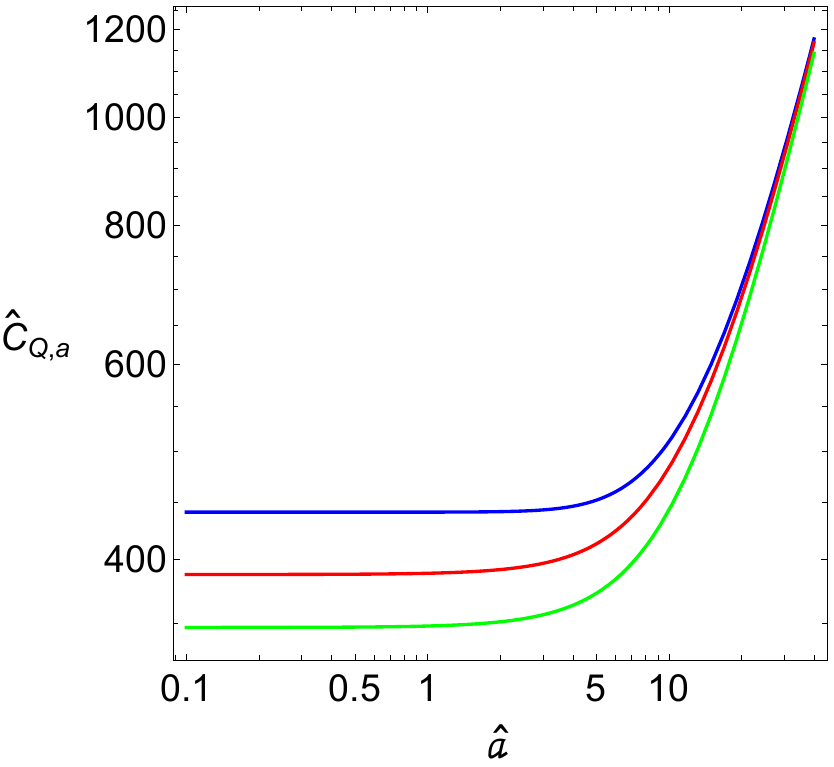}
\hspace{1cm}
\includegraphics[width=6cm]{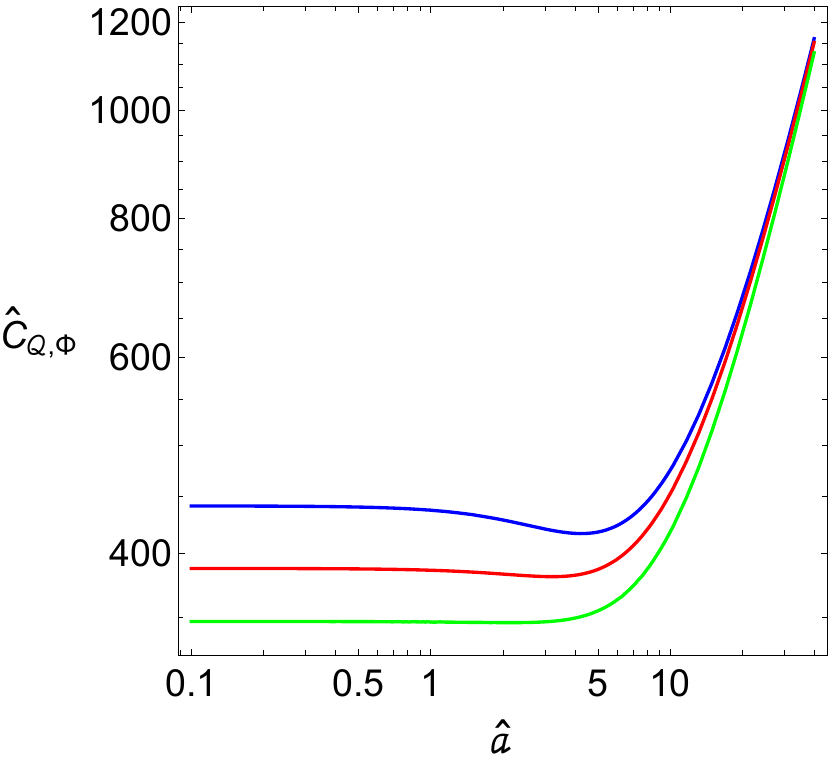}
\caption{{\bf Left Panel:} The dimensionless specific heat $\hat C_{Q,a} = C_{Q,a}/T^2$ as a function of  the dimensionless anisotropy parameter $\hat a = a/T$. {\bf Right Panel:} The dimensionless specific heat $\hat C_{Q,\Phi} = C_{Q,\Phi}/T^2$ as a function of  the dimensionless anisotropy parameter  $\hat a = a/T$. In both panels the blue, red and green solid lines correspond to $\hat Q = Q/T^2$ equal to $0$, $10$ and $20$ respectively.  }  
\label{Fig:PlotcQacQPhivsalpha}
\end{figure}

While the ensemble we consider is for fixed values of the charge density $Q$ 
and anisotropy parameter $a$, the specific heat that is crucial for 
the stability of the solution is a modified specific heat 
(see definition \eqref{specific-heat_Phi}) 
at fixed value of the anisotropisation density
$\Phi$. The stability criteria are analysed in subsection \ref{SubSec:Stab_criteria} and defined by equation \eqref{stability_criteria}.
Since the specific heat at fixed $a$ forms the basis for the modified specific heat, figure \ref{Fig:PlotcQacQPhivsalpha} shows both. In the left panel of figure \ref{Fig:PlotcQacQPhivsalpha}, we plot the dimensionless specific heat $\hat C_{Q,a}$ as a function of $\hat a$, and 
in the right panel we plot the dimensionless specific heat $\hat C_{Q,\Phi}$ 
again as a function of $\hat a$. The plot of the modified specific heat shows that, while non-monotonic with increasing anisotropy, it remains positive for all $\hat a$ values. 
The perturbative expansion of the modified specific heat for small values of $\hat a$
and $\hat Q$, in \eqref{stability_condition_1_smallaQ} of 
appendix \ref{app:SubSec:Stability}, 
is in agreement with the numerical calculation. 
For large values of $\hat a$ we can use the equations \eqref{specific-heat_v1} and \eqref{entropy_largealpha} to find that the specific heat $\hat C_{Q,a}$ scales as $s_{\infty,1} \hat a$, with $s_{\infty,1}$ given in \eqref{entropy_largealpha}.
From figures \ref{Fig:PlotchiPhiTchimuovTvsalpha} and \ref{Fig:PlotcQacQPhivsalpha}, we conclude that the stability criteria hold for all anisotropy values, confirming the background's thermodynamic stability.

\begin{figure}[htb]
\centering
\includegraphics[width=6cm]{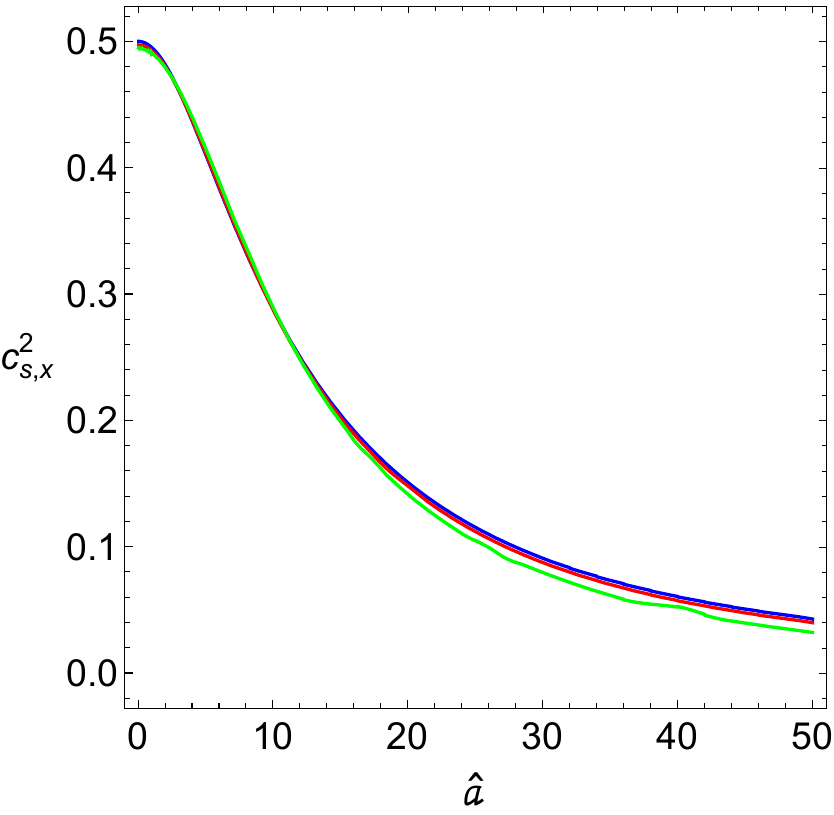}
\caption{The squared speed of sound  in the $x$ direction, i.e. $c_{s,x}^2$,  as a function of  the dimensionless anisotropy parameter $\hat a = a/T$. 
The blue, red and green solid lines correspond to $\hat Q = Q/T^2$ equal to $0$, $10$ and $20$ respectively. 
Note that the blue, red and green curves do not match absolutely perfectly at $\hat a=0$.  This is due to errors that arise in the calculation of second order derivatives of a numerical interpolating function.}  
\label{Fig:Plotcsi2vsalpha}
\end{figure}

In figure \ref{Fig:Plotcsi2vsalpha}, we plot the squared speed of sound along the 
$x$ direction (the speed of sound along the $y$ direction is given by 
\eqref{speedofsound__a_q_fixed}) as a function $\hat a$, for given values of $\hat Q$. 
The plot clearly shows that the speed of sound is almost independent of $\hat Q$. The squared speed of sound as a function of $\hat Q$ (for fixed $\hat a$) forms nearly horizontal lines, providing limited additional insights, so it is excluded from the analysis.

For any $\hat Q$, the squared speed of sound begins at $1/2$ with zero anisotropy and decreases toward zero as $\hat a$ increases. The decrease for small $\hat a$ aligns with the perturbative computation in \eqref{speedofsound_smallq}, and the flow to zero supports the argument in subsection \ref{SubSec:Num_entropy} about the effective reduction of one spatial dimension in the $AdS$ space during the RG flow from small to large distances. 
In contrast, the speed of sound remains nearly constant as $\hat Q$ increases.

The squared speed of sound along the $y$ direction is simply $c_{s,y}^2 = 1 - c_{s,x}^2$, as described in \eqref{speedofsound__a_q_fixed}.  
As the anisotropy parameter increases, the squared speed of sound in the $y$ direction approaches one, while that in the $x$ direction approaches zero.
This might seem counterintuitive, as the axion in \eqref{background_ansatz} is introduced along the $x$ spatial direction.
In a similar scenario in \cite{Ballon-Bayona:2022uyy,DHoker:2009mmn}, where anisotropy arises from a magnetic field along the $z$ direction, the speed of sound increases along the {\it anisotropy}. However, in both cases there
is an RG flow to an $AdS_3$ space where the spatial direction that participates 
(besides the time and the holographic direction), is the one along which the speed 
of sound increases (orthogonal to the anisotropy for the axion and along the {\it anisotropy} for the magnetic field). In our case, this is evident from the analytic solution in equation \eqref{special_2}, while for the magnetic field case, it follows from the RG flow endpoint solution in \cite{Ballon-Bayona:2022uyy,DHoker:2009mmn}. Finally, the $AdS_3$ space at the end of the RG flow exhibits the expected speed of sound, $c^2_{s,y} = 1$, for a conformal fluid in $D=1+1$ dimensions.

\begin{figure}[htb]
\centering
\includegraphics[width=6cm]{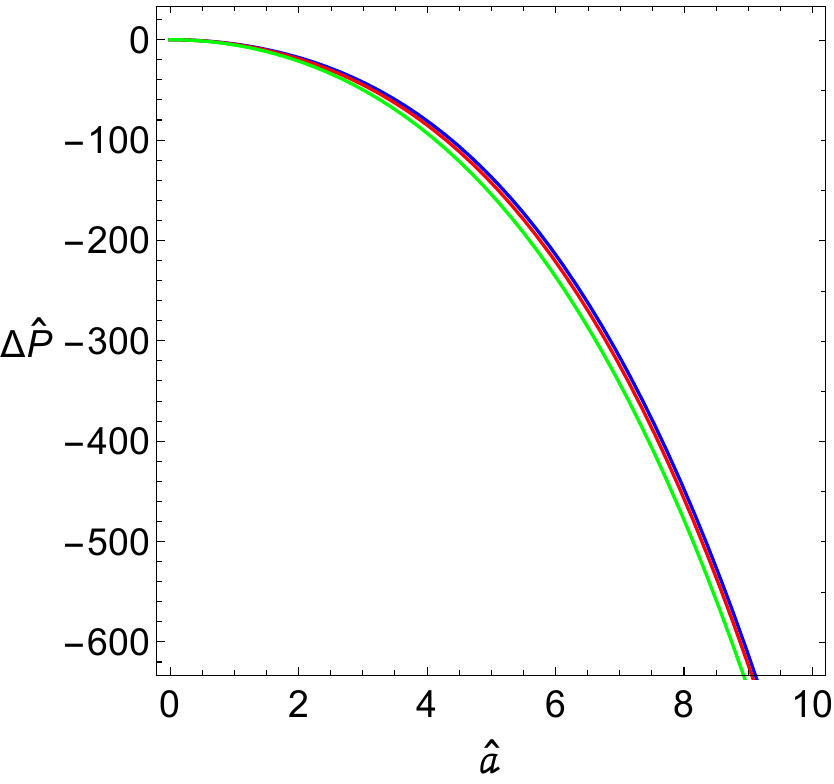}
\caption{The dimensionless pressure anisotropy  $\Delta \hat P = (P_x - P_y)/T^3$ as a function of  the dimensionless anisotropy parameter $\hat a = a/T$.  The blue, red and green solid lines correspond to $\hat Q = Q/T^2$ equal to $0$, $10$ and $20$ respectively. We have numerically verified  that the identity $P_x -P_y = - \Phi a $ holds.  }  
\label{Fig:PlotpxmpyovT3vsalph}
\end{figure}

In figure \ref{Fig:PlotpxmpyovT3vsalph}, we plot the dimensionless pressure anisotropy 
 $\Delta \hat P = (P_x - P_y)/T^3$ as a function of $\hat a$. It takes negative values and it is a 
monotonically decreasing function.  
This indicates that pressure along the anisotropy direction is lower than the orthogonal pressure, with the difference increasing as anisotropy grows. This is in agreement with the observation from figure
\ref{Fig:Plotcsi2vsalpha}, that $c_{s,x}^2<c_{s,y}^2$ in the presence of anisotropy 
and the difference increases as the anisotropy increases. 
Moreover, $\Delta \hat P$ is almost independent of the value
of $\hat Q$. In figure \ref{Fig:PlotpxmpyovT3vsalph} the blue, red and green 
solid lines, corresponding to different values of $\hat Q$, are almost on top of 
each other. 

To construct the plots in figure \ref{Fig:PlotpxmpyovT3vsalph} 
we have used 
the expressions of the pressures coming from the components of the stress tensor in 
\eqref{StressTensor_components}, or equivalently from  \eqref{pressures_action} 
and \eqref{pressure-sum-diff-all-a}. 
We have verified numerically the linear relation between the 
pressure anisotropy and the anisotropisation density that is 
depicted in \eqref{pressures-diff-all-a_v2}.

\subsection{The DC conductivity}

\begin{figure}[htb]
\centering
\includegraphics[width=7cm]{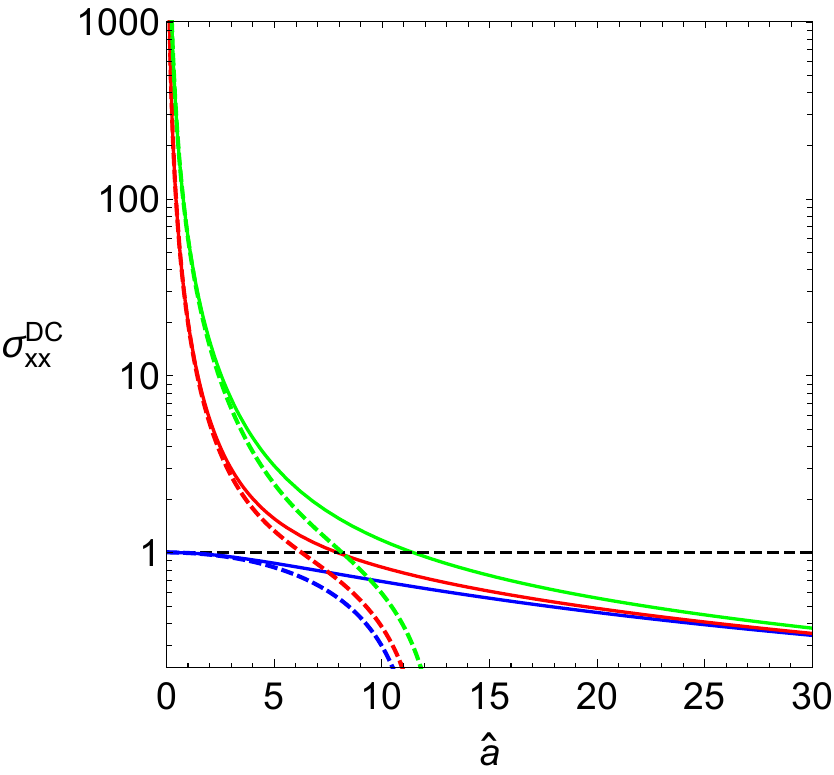}
\caption{The DC conductivity  $\sigma_{xx}^{DC}$ as a function of  the dimensionless anisotropy parameter $\hat a = a/T$. The solid lines represent the numerical solution and the dashed lines the perturbative solution.  The blue, red and green colours correspond to $\hat Q = Q/T^2$ equal to $0$, $10$ and $20$ respectively. The black dashed line represents the limit $\sigma_{xx} \to 1$ when $\hat a \to 0$ in the case of zero charge, i.e. $\hat Q=0$. }  
\label{Fig:Plotconductivityvsalph}
\end{figure}

Figure \ref{Fig:Plotconductivityvsalph} shows the DC conductivity along the $x$ spatial direction as a function of $\hat a$ (note that it is infinite along the $y$ direction since the axion is only along $x$). For small anisotropy, the conductivity diverges, and the numerical results align perfectly with the perturbative result in \eqref{sigma_xx_v2}. For small values of $\hat a$ and $\hat Q$ the perturbative result reduces to \eqref{sigma_xx_largeTexpansion}. 

For large anisotropy, where perturbative expansions fail to provide analytic control, numerical results indicate that conductivity approaches zero for any value of $\hat Q$. 
The fact that the DC conductivity approaches zero in the limit of large anisotropy can be understood from the dictionary in \eqref{sigma_xx_v1}, that expresses this quantity in terms of the horizon data. The first term  is proportional to $\tilde w_{h,0}$ whilst the second term is proportional to $\frac{\hat Q^2}{\hat S^2 \hat a^2}$. Both terms are decreasing functions of $\hat a$, for fixed $\hat Q$, and approach zero in the limit $\hat a \to \infty$, as can be inferred from the right panels of figures \ref{Fig:PlotUh1vsTBQ} and \ref{Fig:Plotsvsq} respectively.
Another interesting explanation for the behaviour of the DC conductivity at large $\hat a$ , particularly for small $\hat Q$, could be drawn from the analytic solution \eqref{special_2}. Although valid for $\hat Q = 0$, it provides a good approximation when the charge is significantly smaller than the anisotropy, the region of interest here. According to this solution, conductivity scales as $1/{\tilde a}$, implying it vanishes in the large-anisotropy limit. Note that the behaviour of the DC conductivity in our model contrasts with the Andrade-Withers isotropic model \cite{Andrade:2013gsa}, where the conductivity approaches one due to two axions linear in spatial directions having equal slopes.

Combining the observations for the conductivity, for finite values of $\hat Q$, 
there is a flow from infinity (small $\hat a$) to zero (large $\hat a$). The vanishing DC conductivity along the $x$ direction in the large-anisotropy limit can be attributed to dimensional reduction. As discussed earlier, for $\hat a \gg \hat Q$, the system flows to a $1+1$ CFT, where only the $y$ and $t$ directions survive. We then expect a vanishing DC conductivity along the $x$ direction as a manifestation of dimensional reduction.

In the case of zero charge, conductivity transitions from unity to zero. 
This behavior is confirmed by both the numerical solution and the perturbative expansion \eqref{sigma_xx_largeTexpansion} for $\hat Q = 0$. 

\begin{figure}[htb]
\centering
\includegraphics[width=7cm]{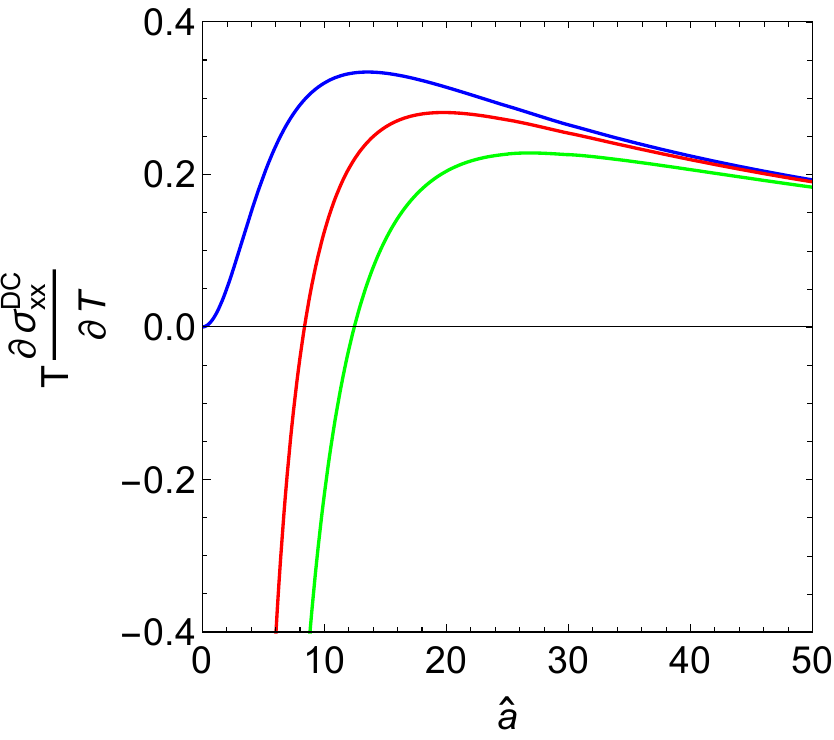}
\hspace{1cm}
\includegraphics[width=6cm]{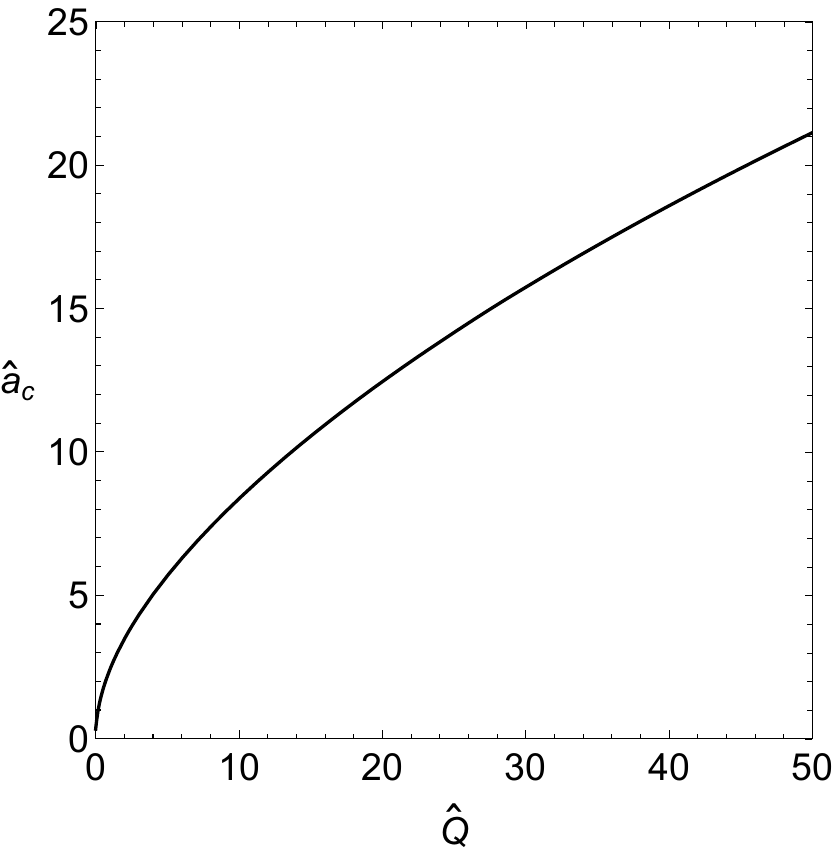}
\caption{{\bf Left Panel:} The derivative of the conductivity with respect to the logarithm of the temperature, i.e. $ T \, \partial \sigma_{xx} / \partial T$,  as a function of  the dimensionless anisotropy parameter $\hat a = a/T$. The blue, red and green colours correspond to $\hat Q = Q/T^2$ equal to $0$, $10$ and $20$ respectively.  {\bf Right Panel:} The critical value $\hat a_c$ where the metal-insulator transition takes place, as a function of $\hat Q = Q/T^2$.}  
\label{Fig:PlotdSigmadLogT}
\end{figure}

We end this section discussing a very interesting transition that occurs in the DC conductivity. On the left panel of figure \ref{Fig:PlotdSigmadLogT} we plot the derivative of the conductivity with respect to the logarithm of the temperature, i.e. $ T \,\partial \sigma_{xx} / \partial T$, which is the appropriate dimensionless quantity in our model. We find that for non-zero $\hat Q$ there are two different regimes depending on the sign of $ \partial \sigma_{xx} / \partial T$ ($T$ is always positive). When the anisotropy parameter is smaller than some critical value, i.e. $\hat a < \hat a_c$, the conductivity increases as the temperature decreases, indicating a metallic behaviour. When the anisotropy parameter is larger than the critical value, i.e. $\hat a > \hat a_c$,  the conductivity decreases as the temperature decreases indicating an insulator behaviour. On the right panel of figure \ref{Fig:PlotdSigmadLogT} we plot the critical value $\hat a_c$ where the metal-insulator transition takes place as a function of $\hat Q$.
Notice that a similar computation for the conductivity has been performed in the case  where anisotropy arises spontaneously from a mechanical deformation \cite{Ji:2022ovs}, and the findings also indicate a metal-insulator transition. However, in the presence of a mechanical deformation the electrical conductivity is a non-diagonal matrix (in our case we consider just the $\sigma_{xx}$ 
entry since the $\sigma_{yy}$ is infinite and the non-diagonal components are zero). In order to see the metal-insulator transition in \cite{Ji:2022ovs}, one needs to diagonalise the matrix and in the new basis  one finds that one direction has a metalic behaviour and the other direction an insulating behaviour. In our scenario the  metal-insulator transition arises along the same direction. Another difference between our model and \cite{Ji:2022ovs} is that in our model the metal-insulator transition takes place without the need of a coupling between the axion and the gauge field \cite{Baggioli:2016oqk}. This coupling is known as charged disorder \cite{Baggioli:2021xuv} and in \cite{Ji:2022ovs}  weak charged disorder is needed.

\section{Conclusions}
\label{Sec:Conclusions}

In this paper we have studied, using the holographic approach, a strongly coupled anisotropic fluid in $2+1$ 
dimensions. The anisotropy is induced through an axion field and the fluid is dual to an asymptotically
$AdS$ black hole, which is a solution of the Einstein-Maxwell-Axion theory in $3+1$ dimensions.  
We began with a hydrodynamic analysis to derive the equation of state and establish criteria for the system's thermodynamic stability. Then we introduced a charged 
anisotropic gravity solution that interpolates between an $AdS_4$ boundary and a black hole 
in the deep IR. A proper holographic renormalisation allowed us to calculate the Gibbs free energy and 
a stress tensor and found that is consistent with the hydrodynamics of a conformal fluid. 
Interestingly, the anisotropy did not disrupt the fluid's initial conformality (prior to axion inclusion). The primary effect of adding the axion is the explicit breaking of translational invariance, leading to finite conductivity.
We investigated this property through holographic computation of the electric DC conductivity.

The equations of motion that are obtained from the Einstein-Maxwell-Axion system can be solved either
perturbatively (for small values of the anisotropy and for finite density) or numerically 
(for generic values of the anisotropy and density). The derivation of the perturbative solution is detailed in 
appendix \ref{app-pert_solution}, and using that solution together with the renormalised action it is possible to compute all the 
relevant thermodynamic quantities (i.e. the temperature, the entropy, the anisotropic enthalpy, 
the pressures and the speeds of sound) and the electric conductivity. Stability criteria are satisfied 
for small values of the anisotropy.

The perturbative solution is analytic but limited to small values of the anisotropy. 
To overcome this limitation we solved numerically the equations of motion and obtain a solution 
that is valid for any value of the anisotropy and the charge. The various thermodynamic quantities, that
we calculate using that solution and the renormalised action,
can be expressed as functions of two dimensionless parameters, namely $\hat Q$ and $\hat a$.
A number of interesting conclusions can be drawn from the plots of the thermodynamic quantities 
as functions of $\hat a$. Here we highlight some of the findings. Notice that we focus on the large 
$\hat a$ behavior, since for small values of $\hat a$  the numerical and the perturbative solutions 
are in agreement and we don't expect {\it the physics} to be very different from  {\it the physics} of $AdS_4$.

When ${\hat a} \gg {\hat Q}$, from the right plot in figure \ref{Fig:Plotsvsq}, we observe that the 
dimensionless entropy density $\hat {\cal S}$ is linear to $\hat a$. From a different approach, it is 
easy to verify (and this is the analysis that we present in appendix  \ref{app:special_solutions})
that there is an analytic solution for zero charge density (this is equivalent to ${\hat a} \gg {\hat Q}$)
of the form $AdS_3 \times R$, where $R$ is a non-compact direction. Calculating $\hat {\cal S}$ 
for this analytic solution we arrive to \eqref{special_2_entropy}, which provides a linear relation between 
$\hat {\cal S}$ and $\hat a$. The slope of the numerical computation is in good agreement with 
the slope from the exact solution (the agreement would be much better if we could expand around the 
linear relation) and the argument is that the numerical solution represents an RG flow between a 
$D = 2 + 1$ CFT at short distances and a $D = 1 + 1$ CFT at long distances from the boundary field 
theory point of view. That RG flow can also be detected through the split between the speeds of 
sound along the $x$ and $y$ directions, as can be seen from figure \ref{Fig:Plotcsi2vsalpha}. 
The $AdS$ space  at the end of the RG flow has the expected speed of sound $c^2_{s,y} = 1$ 
for a conformal fluid in $D=1+1$ dimensions. 
Notice that a different RG flow from the $AdS_4$ boundary to an $AdS_2 \times R^2$ 
horizon region happens  when ${\hat Q} \gg {\hat a}$. This behavior is depicted in the 
left plot of figure \ref{Fig:Plotsvsq}. 

While the stability criteria are satisfied for small values of $\hat a$, to investigate  them 
for generic values of the anisotropy we need the full numerical solution. 
In figures \ref{Fig:PlotchiPhiTchimuovTvsalpha} and 
\ref{Fig:PlotcQacQPhivsalpha} we plotted the ${\hat \chi}_{\mu}$, ${\hat \chi}_{\Phi}$ and 
$\hat C_{Q,\Phi}$ as functions of $\hat a$ and verified that the solution is thermodynamically 
stable, for any value of the anisotropy.

The plot of the DC electric conductivity along the $x$ direction (along the $y$ direction is infinite since 
there is no axion along this direction) was presented in figure \ref{Fig:Plotconductivityvsalph}. An interesting 
observation is that for large values $\hat a$ the conductivity (for any value of $\hat Q$) goes to zero. 
This behaviour is different from the behaviour of the conductivity in the isotropic case 
(i.e. the case with two linear axions with equal slope), where it goes to one \cite{Andrade:2013gsa}. We provided 
a possible explanation, using the analytic solution \eqref{special_2}, where the vanishing DC conductivity along the $x$ direction is a consequence of dimensional reduction. 
It would be interesting to 
understand how this observation is related to the behaviour of the conductivity of anisotropic metals. We also showed at the end of section \ref{Sec:Numerics} that our model leads to a metal-insulator transition driven by the anisotropy and we estimated the critical value of the dimensionless anisotropy parameter $\hat a_c$, where the transition takes place, as a function of the dimensionless charge $\hat Q$, see figure \ref{Fig:PlotdSigmadLogT}.
 
A logical extension of this work would involve adding a second linear axion to the ansatz, with a distinct linear coefficient from the first.
Such a construction, with two axions instead of one, would anisotropise the Andrade/Withers 
solution \cite{Andrade:2013gsa}. This direction would explore the system's behavior when the two slopes are large and differ only slightly. 
We should emphasize here that setting one of the axions to zero (the analysis we present in the 
current paper) means that effectively the slopes of the axions differ significantly. 
This analysis could illuminate an unexplored region of the anisotropic parameter space, potentially uncovering novel physics. Another compelling extension would generalize the Einstein-Maxwell-Axion model to capture transitions between explicit and spontaneous translational symmetry breaking, leading to holographic phonons, as in \cite{Alberte:2017oqx}. 
In the regime of spontaneous breaking of translation symmetry it would be interesting to compare our framework where anisotropy is caused by just one axion field and the stress tensor is diagonal with the framework of \cite{Pan:2021cux, Ji:2022ovs} where anisotropy is caused by two axion fields associated with mechanical deformation and the stess tensor is non-diagonal. 

An alternative direction could be the exploration of the hydrodynamic properties of the system. 
More specifically, it would be interesting to calculate the transport coefficients for perturbations along the 
$xy$ plane in order to study the effect of the anisotropy. 
The standard lore is that the KSS viscosity bound 
\cite{Kovtun:2004de} is violated when there is an explicit breaking of translational invariance\footnote{Breaking of rotational invariance does not necessarily lead to a violation of the KSS bound \cite{Baggioli:2023yvc}.}, 
that in our case is reflected by the presence of the axion, and it would be interesting to investigate the fate of the bound in the current background. 
Recent studies, using a perturbative solution of the equations 
of motion \cite{Li:2024rzd}, show that the KSS bound is violated. 
In light of the metal-insulator transition and the critical line for $\hat a$ that we found for the DC conductivity, it would be interesting to investigate the behaviour of the viscosity for large anisotropy and along the critical line. Another interesting future direction could be the study of the quasinormal modes in the anisotropic background. In the case of the mechanical deformations the analysis of the hydrodynamic properties of the system and of its quasinormal modes 
has been performed in \cite{Xia:2024gba} and \cite{Baggioli:2023dfj}, and it would be interesting to compare with findings from those papers.


\section*{Acknowledgments}

The authors would like to thank Matteo Baggioli and Benjamin Withers for reading our manuscript and for giving very valuable feedback. 
Moreover, the authors thank the anonymous referee for the constructive comments and especially, for calling our attention to the metal-insulator transition, that improved the
quality of the manuscript.
The work of the author A.B-B is partially funded by Conselho Nacional de Desenvolvimento Cient\'\i fico e Tecnol\'ogico (CNPq, Brazil), Grant No. 314000/2021-6, and Coordena\c{c}\~ao de Aperfei\c{c}oamento do Pessoal de N\'ivel Superior (CAPES, Brazil), Finance Code 001. 


\appendix

\section{The perturbative solution for small values of the anisotropy}
\label{app-pert_solution}

The perturbative ansatz we are adopting (we will use the equations in the form they appear in \eqref{system-EOMs-v2}) is 
\begin{equation} \label{perturbative_ansatz}
U\, = \, U_0 +a^2 \, U_2\, , \quad 
V_+ \, = \, 2\, \ln r +a^2 \, V_{+2} \quad \& \quad 
V_-=  a^2 \, V_{-2}
\end{equation}
where $U_0$ is the value of the function $U$ in the absence of anisotropy, i.e. $a=0$, that is written in 
equation  \eqref{HK_solution}. 
From the second equation in \eqref{system-EOMs-v2} we obtain a decoupled equation for $V_{+2}$ 
\begin{equation}
\partial_r\left[ r^2 \, V_{+2}' \right] = 0 \quad \Rightarrow \quad V_{+2}(r) = \frac{c_1}{r} \, . 
\end{equation}
Integrating the differential equation for $V_{+2}$, we obtain 2 integration constants.
The second constant is set to zero, imposing that $V_{+2}$ vanishes at 
infinity.\footnote{Alternatively we could have kept 
that constant and by a redefinition of $x$ and $y$, that constant could be set to zero.} 
Combining the first and the second of the equations in  \eqref{system-EOMs-v2}, 
we obtain a decoupled equation for $V_{-2}$
\begin{equation}
\partial_r\Big[ r^2 \, U_0 \, V_{-2}' \Big] = -\, \frac{1}{2} \, . 
\end{equation}
Imposing that  $V_{-2}$ vanishes at infinity and it is smooth in the IR we arrive to following expression
\begin{equation} \label{Vm2integral}
V_{-2}'(r) = -\, \frac{r-r_h}{2 \, r^2 \, U_0(r)} \quad \Rightarrow \quad 
V_{-2}(r) =  \int_r^{\infty} \frac{y-r_h}{2 \, y^2 \, U_0(y)} \, dy \, . 
\end{equation}
The integral can be calculated analytically and the result is 
\begin{equation} \label{Vm2analytic}
V_{-2}(r) = - \, \frac{1}{2} \sum_{i=1}^3 \frac{\ln \big[r- r_h - x_i\big]}{6\, r_h^2 +8 \, r_h \, x_i +3 \, x_i^2}
\end{equation}
where the three $x_i$'s are the roots of the following cubic equation
\begin{equation} \label{cubic-function}
x^3 \, + \, 4\, r_h \, x^2 \, + \, 6 \, r_h^2 \, x \, + \, \left(3 - \frac{Q^2}{r_h^4}\right) r_h^3 \, = \, 0 \, . 
\end{equation}
Notice that the discriminant of this equation is negative and as a result it has one real root and 
two non-real complex conjugate roots. Nevertheless $V_{-2}$, as it is defined in  \eqref{Vm2analytic}, 
is a real function as long as $Q<\sqrt{3}\, r_h^2$.
Solving  \eqref{cubic-function} and substituting the roots in \eqref{Vm2analytic} we obtain a very complicated expression. 
Expanding for small values of the dimensionless parameter $Q/r_h^2$, 
we obtain an expression of the form $V_{-2}(r) =V_{-2}^{0}(r) + \frac{Q^2}{r_h^4} V_{-2}^{2} (r)$
with the following expression for $V_{-2}^{0}(r)$
\begin{equation} \label{Vm20-expansion}
V_{-2}^{0} = - \, \frac{1}{12\, r_h^2} \Bigg[2\, \sqrt{3} \, \arctan \left[ \frac{\sqrt{3} \, r_h}{ 2\, r + r_h}\right] + 3 
\log \left[\frac{r^2}{r^2 + r\, r_h + r_h^2}\right]\Bigg]
\end{equation}
while the expression for $V_{-2}^{2}$ becomes
\begin{equation} \label{Vm22-expansion}
V_{-2}^{2} = -\,  \frac{1}{18\, r_h^2} \Bigg[2\, \sqrt{3} \, \arctan \left[ \frac{\sqrt{3} \, r_h}{ 2\, r + r_h}\right] -9 \log \left[\frac{r^2}{r^2 + r\, r_h + r_h^2}\right] -\frac{3 \left(2 \, r_h+r\right) r_h}{r^2 + r\, r_h + r_h^2}-\frac{9 \, r_h}{r}\Bigg] \, . 
\end{equation}
The value of the constant $c_1$ is fixed by requiring that the correction to $V$ vanishes at $r=r_h$. 
In fact this is the way we fix it in the numerical solution. That means 
\begin{equation}
V_{+2}(r_h) + V_{-2}(r_h) = 0 \quad \Rightarrow \quad c_1 =  - \, V_{-2}(r_h) \, r_h \, . 
\end{equation}
Expanding \eqref{Vm2analytic} we can calculate a perturbative expression for $c_1$
\begin{eqnarray} \label{c1-pert}
c_1 &=& \frac{1}{r_h} \Bigg[c_{10} + c_{12} \,  \frac{Q^2}{r_h^4} + c_{14}\,  \frac{Q^4}{r_h^8} +c_{16} \,  \frac{Q^6}{r_h^{12}} + c_{18}  \, \frac{Q^8}{r_h^{16}}  + {\cal O} \left(\frac{Q^{10}}{r_h^{20}} \right)\Bigg] 
\end{eqnarray}
where the components of the expansion are
\begin{eqnarray} \label{c1-components}
&& c_{10}= \frac{\sqrt{3} \, \pi -9 \log 3}{36} \, , \quad
 c_{12}=  \frac{\sqrt{3} \, \pi +9 \left(3 \log 3 -4\right)}{54} \, , \quad 
 c_{14}= \frac{192-19 \sqrt{3} \, \pi -81 \log 3}{108}
\nonumber \\
&& 
 c_{16}= \frac{140 \sqrt{3} \, \pi -762}{243}  \quad \& \quad
 c_{18}= -5\, \frac{172 \sqrt{3} \, \pi -27 \left(5+27 \log 3\right)}{729} \, . 
\end{eqnarray}
The equation for $U_2$ is the following
\begin{equation}
\partial_r\Big[ r^2 \, U_2' \Big] = \frac{c_1\, r_h^4}{r^3} \, 
\left[\left(1+\frac{Q^2}{r_h^4}\right) \, \frac{r}{r_h}-
\frac{6\, Q^2}{r_h^4}+\frac{2\, r^4}{r_h^4}\right] \, . 
\end{equation}
Integrating the equation of motion for $U_2$ and imposing the requirement that $U_2(r_h)=0$ fixes one of the integration constants. The second is fixed by using the constraint. The  expression for $U_2$ that we obtain is
\begin{equation}
U_2 (r) = \left(\frac{r}{r_h}-1\right) \frac{r_h^3}{r^3}  \,\Bigg\{r_h \, c_1 \left[\frac{Q^2}{2\, r_h^4} 
\left(2+ \frac{r}{r_h}\right)+\frac{r^2}{r_h^2} +\frac{r^3}{r_h^3} - \frac{r}{2 \, r_h} \right] 
- \frac{r^2}{4\, r_h^2} \Bigg\} \, . 
\end{equation}
Expanding the above expressions close to the boundary and close to the horizon we obtain the following 
values for the asymptotic expansion parameters, that we also list in the main text in equation \eqref{Pert-solution_parameters}, 
\begin{eqnarray} 
&&\frac{U_{\infty, 1}}{a^2 \, r_h}= -\, V_{-2}(r_h)\, , \quad 
\frac{U_{\infty, 3}}{r_h^3} = - \, 1 - \frac{Q^2}{r_h^4} + \frac{a^2}{4 \, r_h^2}
\Bigg[1+2\, r_h^2 \left(3-\frac{Q^2}{r_h^4} \right)V_{-2} (r_h)\Bigg]
\nonumber\\[5pt]
&& 
\frac{v_{\infty, 3}}{a^2\, r_h} =- \,  \frac{1}{12}  \, , \quad 
\frac{v_{h,0}}{r_h}=1 \, , \quad 
\frac{\mu \, r_h}{Q} \, =\, 1+ \frac{\alpha^2}{2} V_{-2} (r_h)
\\[5pt]
&& 
\frac{w_{h,0}}{r_h} =1 -  a^2 \, V_{-2}(r_h) \quad \& \quad 
\frac{U_{h,1}}{r_h} = 3 - \frac{Q^2}{r_h^4} 
- \frac{a^2}{4 \, r_h^2}
\Bigg[1+6\, r_h^2 \left(1+\frac{Q^2}{r_h^4} \right)V_{-2} (r_h)\Bigg]\, .
\nonumber
\end{eqnarray}
These are the values we are going to use in expanding all the physical quantities around the isotropic HK solution. 


\subsection{Intermediate expressions}

In this subsection of the appendix we are gathering intermediate expressions 
for several physical quantities. In the beginning we are listing expansions 
around the Harnoll-Kovtun
solution (see \cite{Hartnoll:2007ai}), for small value of the parameter $\tilde a$.
When those expressions combine with the expansion of the temperature in 
terms of $\tilde a$ (as it appears in equation \eqref{T+S_small_tilde_a}), 
we are able to obtain the expansions in terms of $\hat a$ and $\hat Q$.
In the second part, we are listing expressions of quantities in terms of 
the dimension-less function $\hat {\cal F}$ and its derivatives to $\hat a$ and  ${\hat {\cal Q}}$, that are useful in the calculation of the speed of sound.

Substituting the values of the parameters from 
\eqref{Pert-solution_parameters} into  \eqref{StressTensor_components},
we obtain the following expressions for the different components of 
the energy-momentum tensor in the limit of small $\tilde a$
\begin{eqnarray}  \label{StressTensor_components_v2}
\frac{ {\cal E}}{2\,\sigma \, r_h^3}&= &1 + {\tilde Q}^2 - \frac{{\tilde a}^2}{4}
\Bigg[1+2 \left(3-{\tilde Q}^2 \right)V_{-2}({\tilde Q}) \Bigg] 
\nonumber \\
\frac{P_x}{\sigma \, r_h^3}&= &1 + {\tilde Q}^2 -  \frac{{\tilde a}^2}{4}
\Bigg[3+2  \left(3-{\tilde Q}^2\right) V_{-2}({\tilde Q}) \Bigg] 
\\
\frac{P_y}{\sigma \, r_h^3}&= &1 + {\tilde Q}^2+  \frac{{\tilde a}^2}{4}
\Bigg[1-2  \left(3- {\tilde Q}^2 \right) V_{-2}({\tilde Q})\Bigg] \, . 
\nonumber
\end{eqnarray}
The expansions of $\hat S$, $\hat Q$, $\hat \Phi$ and
$\hat a$  (as they are defined in \eqref{dim-less-Q-S-a-Phi} and after 
using \eqref{T+S_small_tilde_a}), 
in the limit of small $\tilde a$ are
\begin{eqnarray} \label{dimless_ratios_all_Q}
{\hat {\cal S}}&=& \frac{64\, \sigma \, \pi^3}{(3-{\tilde Q}^2)^2} 
\Bigg\{1+ \frac{1}{2}\, \frac{{\tilde a}^2}{3-{\tilde Q}^2}
\Big[1+ 8\, {\tilde Q}^2 \, V_{-2} ({\tilde Q})\Big]\Bigg\} 
\, , \quad 
{\hat \Phi}= \frac{16\, \pi^2\, \sigma \, {\tilde a}}{(3-{\tilde Q}^2)^2}
\nonumber \\
{\hat Q} &=& \frac{4^2 \, \pi^2 \, {\tilde Q}}{(3-{\tilde Q}^2)^2}\Bigg\{1+ \frac{{\tilde a}^2}{3-{\tilde Q}^2}
\Big[1+ 6\, (1+{\tilde Q}^2) \, V_{-2}({\tilde Q})\Big]\Bigg\}^{1/2} \\ 
{\hat a}&=& \frac{4\, \pi \, {\tilde a}}{3-{\tilde Q}^2}\Bigg\{1+ \frac{1}{4} \, \frac{{\tilde a}^2}{3-{\tilde Q}^2}
\Big[1+ 6\, (1+{\tilde Q}^2 ) \, V_{-2}({\tilde Q})\Big]\Bigg\} \, .  \nonumber
\end{eqnarray}
The expression for the dimensionless function ${\hat {\cal F}}$ 
(see the definition in  \eqref{def_dimless_free_energy})
in terms of the dimensionless quantities $\tilde Q$ and $\tilde a$ reads
\begin{equation} 
{\hat {\cal F}} =  192 \, \sigma \, \pi^3 \, \frac{{\tilde Q}^2 -\frac{1}{3}}{(3-{\tilde Q}^2)^3} 
\Bigg\{1+ \frac{5}{6}\, {\tilde a}^2  \frac{{\tilde Q}^2-\frac{3}{5}}{(3-{\tilde Q}^2)\big({\tilde Q}^2-\frac{1}{3}\big)}
\Big[1+ \frac{24}{5}\, \frac{{\tilde Q}^2({\tilde Q}^2+1)}{ {\tilde Q}^2-\frac{3}{5} } \, V_{-2} ({\tilde Q})\Big]\Bigg\} \, . 
\label{dimless_free_energy_v1}
\end{equation}

The specific heat (an essential ingredient in the calculation of the speed 
of sound) can be expressed in terms of the dimension-less function 
$\hat {\cal F}$ and its derivatives to $\hat a$ and  ${\hat {\cal Q}}$ as follows
\begin{equation} \label{specific_heat_v2}
\frac{C_{a, {\cal Q}}}{T^2} = - 6\, {\hat {\cal F}} -4 \, {\hat {\cal Q}} \,{\hat a}  \,   \partial_ {{\hat {\cal Q}}} \partial_ {{\hat a} } {\hat {\cal F}} 
+ 4\, {\hat a} \, \partial_ {\hat a}  {\hat {\cal F}} +  6\, {\hat {\cal Q}} \, \partial_ {\hat {\cal Q}}  {\hat {\cal F}}  -
{\hat a}^2 \, \partial^2_ {\hat a}  {\hat {\cal F}} -  4\, {\hat {\cal Q}}^2 \, \partial^2_ {\hat {\cal Q}}  {\hat {\cal F}} \, . 
\end{equation}
The pyro-charge and the pyro-anisotropic coefficients in terms of $\hat {\cal F}$ read 
\begin{equation} \label{pyro_charge_anisotropic_v2}
\frac{\xi_{\Phi}}{T} = - \, 2\, \partial_{\hat a} {\hat {\cal F}}  +  
{\hat a} \, \partial^2_{\hat a} {\hat {\cal F}}  + 
 2\, {\hat {\cal Q}} \, \partial_{{\hat a}} \partial_{\hat {\cal Q}} {\hat {\cal F}} 
 \quad \& \quad 
 \xi_{\mu}  = \partial_ {\hat {\cal Q}}  {\hat {\cal F}}  -  
 2\,  {\hat {\cal Q}}  \, \partial^2_{ \hat {\cal Q}} {\hat {\cal F}}   - 
 {\hat a}\, \partial_{\hat a} \partial_{\hat {\cal Q}}  {\hat {\cal F}}\, . 
\end{equation}
Combining the entropy density from \eqref{dim-less-quantities}, together with the 
pyro-charge and the pyro-anisotropic coefficients from 
\eqref{pyro_charge_anisotropic_v2}, we arrive to the following expression
\begin{equation} \label{pressure_x_deriv}
\frac{{\cal S}+ {\cal Q}\, \xi_{\mu} - a \, \xi_{\Phi}}{T^2} = 
- \, 3\, {\hat {\cal F}}  - 
3\, {\hat {\cal Q}} \, {\hat a} \, \partial_ {{\hat a}} \partial_ {\cal Q} {\hat  {\cal F}}  + 
3\,  {\hat a}  \, \partial_ {\hat a}  {\hat  {\cal F}}  
+  3 \,  {\cal Q} \, \partial_{ {\cal Q}} {\hat  {\cal F}}   -  
2\,  {\cal Q}^2 \, \partial^2_{ {\cal Q}} {\hat  {\cal F}}   -  
{\hat a} ^2 \, \partial^2_ {\hat a}  {\hat  {\cal F}} \, . 
\end{equation}
%


\subsection{Stability conditions for small values of $\hat Q$}
\label{app:SubSec:Stability}

In this subsection of the appendix we are listing the expressions of the 
specific heat at fixed $\Phi$ (namely $C_{{\cal Q}, \Phi}$) as well as of 
the charge (namely $\chi_{\mu}$) and of the 
anisotropic (namely $\chi_{\phi}$) susceptibility for small values of 
$\hat a$ and $\hat Q$. To do so we are using the form of the free energy from
\eqref{dimless_free_energy_v2} together with the expressions 
\eqref{charge_anisotropic_susc}, \eqref{specific-heat_Phi}, 
\eqref{specific_heat_v2} and \eqref{pyro_charge_anisotropic_v2}. For the 
specific heat we have
\begin{eqnarray} \label{stability_condition_1_smallaQ}
&& \frac{{\hat C}_{{\cal Q},\Phi}}{\sigma} =  
\Bigg[ \frac{128 \pi ^3}{9}-\frac{3}{\pi }\, \hat{Q}^2+\frac{1215}{512 \,  \pi^5}  \, \hat{Q}^4-
\frac{32805}{16384 \, \pi ^9} \, \hat{Q}^6+ \cdots \Bigg]
\nonumber \\[5pt]
&& \qquad \qquad +\, \hat{a}^2
\Bigg[ -\frac{4 \pi }{3}+\frac{117}{64 \, \pi ^3}  \, \hat{Q}^2-\frac{35235}{16384 \, \pi
   ^7}  \, \hat{Q}^4 +\frac{10123623}{4194304 \, \pi ^{11}}  \, \hat{Q}^6+ \cdots \Bigg] \, . 
\end{eqnarray}
For the charge susceptibility the expression becomes
\begin{eqnarray} \label{stability_condition_2_smallaQ}
\sigma \, {\hat \chi}_{\mu}   & = & 
\Bigg[\frac{3}{16 \pi }-\frac{243}{4096 \, \pi ^5}  \, \hat{Q}^2 +\frac{10935}{262144 \, \pi^9}  \, \hat{Q}^4 + \cdots \Bigg]  
\nonumber \\[5pt]
&& \qquad  - \, 
   \hat{a}^2  \Bigg[ \frac{9}{1024 \, \pi ^3}- \frac{2187}{131072 \, \pi ^7}  \, \hat{Q}^2  + 
   \frac{1476225}{67108864 \, \pi^{11}}  \, \hat{Q}^4 + \cdots \Bigg] \, .
\end{eqnarray}
For the anisotroppic susceptibility the expression reads
\begin{equation} \label{stability_condition_3_smallaQ}
\frac{\hat {\chi}_{\Phi}}{\sigma}  =
 \frac{4 \, \pi }{3} + \frac{9}{64 \, \pi ^3}\, \hat{Q}^2- \frac{729}{16384 \,  \pi ^7} \, \hat{Q}^4 + 
 \frac{98415}{4194304 \, \pi ^{11}} \, \hat{Q}^6+ \cdots \, . 
\end{equation}
Those expressions are compatible with the plots from the figures 
\ref{Fig:PlotchiPhiTchimuovTvsalpha} and
\ref{Fig:PlotcQacQPhivsalpha} and provide an analytic proof for the 
stability of the solution in the range of small $\hat Q$ and small $\hat a$.


\subsection{Expressions in the limit of zero temperature}
\label{app:zeroT}

In this subsection of the appendix we are listing expressions 
for several physical quantities in the limit of zero temperature, 
when the anisotropy is kept small. Using the expansion of the temperature from 
\eqref{T+S_small_tilde_a} to solve the equation $T=0$ perturbatively, 
it is possible to determine the relation between the charge and the anisotropy 
in the limit of zero temperature. It reads
\begin{equation} \label{zeroT_q_expansion}
{\tilde Q} = \sqrt{3} - \frac{1}{4}\, {\tilde a}^2 
\Bigg[1 - \frac{2}{\sqrt{3}}\, \log {\tilde a}\Bigg] + \cdots  \, . 
\end{equation}
Notice that for zero anisotropy the limit of zero temperature corresponds to 
$\tilde Q$ close to $\sqrt{3}$. 

Expanding \eqref{dimless_ratios_all_Q} around the value of $\tilde Q$ that 
we calculated in \eqref{zeroT_q_expansion} and for small $\tilde a$,
we obtain the following expression for the dimensionless ratio of the entropy 
\begin{equation} \label{SoverT2_smallTexpansion_v2}
{\hat {\cal S}} =   {\hat {\cal S}}\big\vert_{\hat a \,=\,0}  + \,  \frac{\sigma \, \pi}{6}  \, {\hat a}^2 
\Bigg[ 1+ 3^{1/4} \pi \frac{1}{\hat Q^{1/2}} +\frac{2 \pi ^2}{3^{1/2}}  \frac{1}{\hat Q} + 
\frac{5 \pi ^3}{2 \times 3^{5/4}} \frac{1}{\hat Q^{3/2}} \cdots \Bigg]
\end{equation}
with 
\begin{equation} \label{SoverT2_smallTexpansion_v1}
 {\hat {\cal S}}\big\vert_{\hat a \,=\,0}  =  \frac{4\, \sigma \, \pi}{\sqrt{3}} \, {\hat Q} \Bigg[ 1+\frac{2 \, \pi }{3^{3/4}} 
 \frac{1}{\hat Q^{1/2}} +\frac{4 \pi ^2}{3^{3/2}}  \frac{1}{\hat Q} +\frac{7 \pi ^3}{3^{9/4}}  \frac{1}{\hat Q^{3/2}} \cdots \Bigg] \, . 
\end{equation}
For the anisotropisation density $\Phi$ the corresponding expressions become
\begin{equation} \label{PhioverT2_smallTexpansion_v1}
{\hat \Phi} = \frac{\sigma}{3^{1/4}} \, {\hat a} \, {\hat Q^{1/2}} \Bigg[ 1+ \frac{\pi}{3^{3/4}} \frac{1}{\hat Q^{1/2}} +
\frac{\pi ^2}{2 \times 3^{1/2}}  \frac{1}{\hat Q} + 
\frac{2 \,  \pi ^3}{3^{9/4}} \frac{1}{\hat Q^{3/2}} \cdots \Bigg] \, . 
\end{equation}
Expanding \eqref{dimless_free_energy_v1}, we arrive to the following expression of the 
dimensionless function ${\hat {\cal F}} $
\begin{eqnarray}  \label{dimless_free_energy_v3}
{\hat {\cal F}} &= &  \frac{8\, \sigma}{3^{3/4}} \, \hat{Q}^{3/2} \Bigg[ 1 
-\frac{3^{1/4} \, \pi }{2} \frac{1}{\hat{Q}^{1/2}} -\frac{\pi ^2}{2 \, \sqrt{3}}  \frac{1}{\hat{Q}} 
-\frac{2 \, \pi ^3}{3^{9/4}} \frac{1}{\hat{Q}^{3/2}} -\frac{7 \, \pi^4}{72}  \frac{1}{\hat{Q}^{2}} \Bigg] 
\nonumber \\
\quad \quad && 
 -\frac{\sigma}{2 \times 3^{1/4}} {\hat a}^2 \, {\hat{Q}}^{1/2} \Bigg[1+\frac{\pi }{3^{3/4} } \frac{1}{\hat{Q}^{1/2}} 
+\frac{\pi ^2}{2 \, \sqrt{3}}  \frac{1}{\hat{Q}} + \frac{2 \, \pi ^3}{3^{9/4}} \frac{1}{\hat{Q}^{3/2}} 
+ \frac{5 \, \pi^4}{72}  \frac{1}{\hat{Q}^{2}} \Bigg] \, . 
\end{eqnarray}
Combining \eqref{specific_heat_v2}, \eqref{pressure_x_deriv} and 
\eqref{dimless_free_energy_v3}, we arrive to the following expression for  
the speed of sound along the $x$ direcion that is valid for small $\hat a$ 
and large $\hat Q$
\begin{equation} \label{speedofsound_largeq}
c_{s,x}^2=  \frac{1}{2} -\left[\frac{\sqrt[4]{3}}{16 \pi}\, \sqrt{\frac{1}{\hat{Q}} } - 
\frac{1}{16 \, \sqrt{3}}\, \frac{1}{\hat{Q}}-\frac{\pi }{96 \sqrt[4]{3}}\, \frac{1}{\hat{Q}^{3/2}} +
\frac{101 \pi ^3}{576\times 3^{3/4}} \, \frac{1}{\hat{Q}^{5/2}} + \cdots \right] 
{\hat a}^2 \, . 
\end{equation}
The conductivity at zero temperature comes from the expansion of \eqref{sigma_xx_v2}
\begin{equation} \label{sigma_xx_smallTexpansion}
\sigma_{xx} =   \frac{1}{{\hat a}^2}
\Bigg[4 \, \sqrt{3}\,  {\hat Q} - \frac{8 \, \pi}{\sqrt[4]{3}} {\hat Q}^{1/2}+ 
\frac{4 \, \pi ^3}{3^{7/4}} \, \frac{1}{{\hat Q}^{1/2}} \Bigg] +\frac{1}{2}+ \frac{\pi }{2\times  3^{3/4}} \, 
   \frac{1}{{\hat Q}^{1/2}}+ \cdots \, . 
\end{equation}


\section{Analytical solutions (large anisotropy \& large charge)}
\label{app:special_solutions}

It is possible to obtain special analytic solutions to the equations of motion \eqref{system-EOMs}, which could be useful
for the field theory interpretation. 

Setting to zero the anisotropy parameter $a$, it is possible to solve the equations of motion in \eqref{system-EOMs}
under the assumption $W(r)=V(r)$ (a natural requirement since there is no anisotropy). The solution is 
\begin{equation} \label{special_1}
ds^2 =   - 6\left(r^2 -r_h^2\right) \, dt^2  + \frac{dr^2}{6\left(r^2 -r_h^2\right)} + \frac{Q}{\sqrt{3}}\left( d x^2 + d y^2 \right)
\end{equation}
with
\begin{equation}
 f = \sqrt{3} \left(r-r_h\right) \, . 
\end{equation}
The temperature and the entropy read
\begin{equation} \label{special_1_entropy}
T = \frac{3 \, r_h}{\pi} \quad \& \quad {\cal S} = \frac{4}{\sqrt{3}} \, \sigma \, \pi \, Q \quad \Rightarrow \quad 
{\hat {\cal S}} = \frac{4\, \sigma \, \pi }{\sqrt{3}} \, {\hat Q} \, . 
\end{equation}
Notice that \eqref{special_1_entropy} corresponds to the first term in the small temperature expansion of \eqref{SoverT2_smallTexpansion_v1}.


Setting to zero the charge density (i.e. $Q=0$), it is possible to 
solve analytically the equations of 
motion in \eqref{system-EOMs}. The solution is 
\begin{equation}  \label{special_2}
ds^2 =   - \frac{3}{2}\left(r^2 -r_h^2\right) \, dt^2  + \frac{dr^2}{\frac{3}{2}\left(r^2 -r_h^2\right)} + \frac{a^2}{6} \,  d x^2 + r^2 \, d y^2
\end{equation}
The temperature and the entropy read
\begin{equation}
T = \frac{3 \, r_h}{4\, \pi} \quad \& \quad {\cal S} = 4 \, \sigma \, \pi \, r_h \, \frac{a}{\sqrt{6}} \, . 
\end{equation}
The expression of the entropy density as a function of the anisotropy parameter and the temperature reads
\begin{equation}  \label{special_2_entropy}
{\cal S} =  \frac{16\, \sigma \, \pi^2}{3\, \sqrt{6}} \, \alpha \, T \quad \Rightarrow \quad  
{\hat {\cal S}} =   \frac{16\, \sigma \, \pi^2}{3\, \sqrt{6}} \, {\hat a}\, . 
\end{equation}
This is the prediction for the behavior of the entropy density for large values of the anisotropic parameter 
(with respect to the density), which is verified by the numerical computation.

In principle it should be possible to expand around the two special 
solutions in \eqref{special_1} and \eqref{special_2} and obtain more terms in the expression of the 
dimension-less entropy with respect to the dimension-less ratios 
$\hat a$ and $\hat Q$. In practice this is 
not possible, since those two solutions do not have an $AdS_5$ nature and it is not clear how to impose the 
boundary conditions in the UV. 

The analytic solutions in \eqref{special_1} and \eqref{special_2} can be seen as 
the limit of the full numerical solution for ``large" values of the charge or the 
anisotropy parameter. 
From a field theory perspective this is equivalent to the small temperature limit and 
an alternative expansion could be around the zero temperature. While for zero 
anisotropy such a limit exists, for zero charge (in other words expanding the 
entropy around \eqref{special_2_entropy}) this is not possible. 
This can be seen by focusing on the first 
and fourth equations of  \eqref{system-EOMs-v2}. If one sets $U(r_h)=U'(r_h)=0$, 
then the system 
is consistent only if the anisotropy is set to zero. 
Since the temperature cannot be set to zero, this means that the expansion has to 
be performed around a finite and large value of the anisotropy (so $\hat a$ is large). In this case the equations do not decouple and the system cannot be solved analytically.


\section{Calculation of the conductivity}
\label{app-conductivity}

In order to be self contained, in this appendix, we reproduce the calculation of the conductivity, 
following \cite{Donos:2014cya} (see also \cite{Blake:2015ina}).
To be more generic, we keep the setting with two axions of that reference, i.e  $\chi_1 = a_1 \, x$  
and $\chi_2 = a_2 \, y$, however in the end we will set the second axion to zero, i.e. $a_2 =0$. 
In this way the $\sigma_{yy}$ component of the conductivity will be infinite, 
a known property in the presence of translational invariance.

To perturb the action \eqref{4d-action} we use the following conventions
\begin{equation}
g_{\mu \nu} \rightarrow g_{\mu \nu} + h_{\mu \nu} 
\,\,\,\, {\rm with}  \,\,\,\,
\delta g^{\mu \nu}   \, = \, - \,  g^{\mu \rho} \,  h_{\rho \lambda} \,  g^{\lambda \nu} 
 \,\,\,\, \&  \,\,\,\, 
 \delta g = -g \, g_{\mu \nu} \, \delta g^{\mu \nu} =  g \, g^{\mu \nu} \, h_{\mu \nu} \, . 
\end{equation}
To calculate the electric conductivity along the directions $x$ and $y$, we turn on a constant electric field along those 
directions with magnitudes $E_x$ and $E_y$, respectively. 
That in turn will generate electric currents in the  $x$ and $y$ directions, 
which we will label ${\cal J}^x$ and ${\cal J}^y$. 
The expressions for the conductivities (more precisely $\sigma_{xx}$ and $\sigma_{yy}$) will be obtained 
once we have calculated the expressions for ${\cal J}^x$ and ${\cal J}^y$, in terms of horizon data. 
We consider the following small perturbation about the black hole solutions
\begin{eqnarray} \label{pert_ansatz} 
& \delta A_{x} = -E_x \, t + a_x(r) \quad \& \quad \delta A_{y} = -E_y \, t + a_y(r) &
\nonumber \\[5pt]
& h_{t x} =g_{x x }\, H_{t x}(r) \quad \& \quad h_{t y} =g_{yy }\, H_{t y}(r) &
\nonumber \\[5pt]
& h_{r x} =g_{x x }\, H_{r x}(r) \quad \& \quad h_{r y} =g_{yy }\, H_{r y}(r) &
\end{eqnarray}
and all the other components of the perturbation of the metric and of the gauge field are set to zero. 
Lets emphasize at this point that this is consistent with the equations of motion for the fluctuations.  
We start the analysis with the fluctuation of the Maxwell equation from  \eqref{Maxwell_axion_EOM}
\begin{equation}  \label{Maxwell_v2}
\partial_{\mu} \left[\sqrt{-g} \, g^{\mu \alpha } \, g^{\nu \beta }  \, F_{\alpha \beta} \right] = 0 \, . 
\end{equation}
Applying \eqref{pert_ansatz} in  \eqref{Maxwell_v2} we arrive to following expression
\begin{equation}  \label{Maxwell_perturbation}
\partial_{\mu} \Bigg[\sqrt{-g} \, \left[\frac{1}{2} \, h^{\lambda}{}_{ \lambda}   \, F^{\mu \nu} 
- h^{\mu}{}_{ \lambda} \,F^{\lambda \nu} - h^{\nu}{}_{ \lambda} \,F^{\mu \lambda } 
+ \delta F^{\mu \nu} \right] \Bigg] \, = \, 0 \, . 
\end{equation}
The only non-trivial components of that equation are the $x$ and $y$ and can be written in the form 
$\partial_r {\cal J}^x=\partial_r {\cal J}^y = 0 $. The expressions for ${\cal J}^x$ and ${\cal J}^y$, after using the 
ansatz \eqref{background_ansatz}, become\footnote{The relation between the 
currents that we define in \eqref{Jx-Jy_v2} and the currents 
in \cite{Donos:2014cya} is ${\cal J}_{BDL}^{(x,y)}= 4 \,  {\cal J}_{here}^{(x,y)}$. 
The reason behind this redefinition is the following: In one considers the isotropic case (i.e. $a_1=a_2=a$ and  $W(r)=V(r)$) and the limit of large anisotropy, the conductivity flows to the unity. If we don't consider this redefinition the conductivity would flow to 4.} 
\begin{equation} \label{Jx-Jy_v1}
 {\cal J}^x = - e^{W - V} \, U \, a_x' - e^{V+ W} \, f' \, H_{t x} \quad \& \quad  
 {\cal J}^y  = - e^{V - W} \, U \, a_y' - e^{V+ W} \, f' \, H_{t y}
\end{equation}
where the right-hand side can be evaluated at any value of the radial distance, including the horizon of the 
black hole.
Perturbing the Einstein equation in \eqref{einstein_v2} we arrive to the following fluctuation equation
\begin{equation}
\delta R_{\mu \nu} +3 \, \delta g_{\mu \nu }  =  2 \, \delta T^{A}_{\mu \nu} + 
\frac{1}{2} \sum_{i=1}^2 \left(\partial_{\mu} \delta \chi_i \, \partial_{\nu} \chi_i +
\partial_{\mu} \chi_i \, \partial_{\nu} \delta \chi_i\right) \,
\end{equation}
where we perturb the 2 axions as $\chi_i \rightarrow \chi_i + \delta \chi_i(r)$ for $i=1,2$. 
The perturbation of the Ricci tensor and of $T^{A}_{\mu \nu}$ are given by the following expressions
\begin{equation}
\delta R_{\mu \nu}\,=\,
\frac{1}{2}\,\Bigg[
D_{\rho}\,D_{\mu}\,h^{\rho}_{\,\,\nu}\,+\,
D_\rho\,D_\nu\,h^{\rho}_{\,\,\mu}\,-\,
D_\rho\,D^\rho\,h_{\mu \nu}\,-\,
D_\mu\,D^\nu\,h^\rho_{\,\, \rho}\Bigg]
\end{equation}
and
\begin{equation}
\delta T^{A}_{\mu \nu} =  F_{\mu \rho} \, \delta F_{\nu}{}^{\rho} +  F_{\nu \rho} \, \delta F_{\mu}{}^{\rho} 
- F_{\mu}{}^{\alpha} \, F_{\nu}{}^{\beta} h_{\alpha \beta} 
- \frac{ h_{\mu \nu}}{4}\, F^2
- \frac{g_{\mu \nu}}{2} \Big[\delta F_{\alpha \beta} \, F^{\alpha \beta} 
- h^{\alpha}{}_{\rho} F^{\rho \beta} \, F_{\alpha \beta}\Big] .
\end{equation}
We find two equations which can be solved algebraically it terms of $\delta \chi_1' $ and $\delta \chi_2'$, giving
\begin{equation} \label{delta_chi_prime}
\delta \chi_1' = a_1 \, H_{xr} +\frac{4\, E_x \, Q}{a_1 \, U}  \, e^{- V - W} \quad \& \quad 
\delta \chi_2' = a_2 \, H_{yr} +\frac{4\, E_y \, Q}{a_2 \, U}  \, e^{- V - W} 
\end{equation}
as well as another two coupled second order differential equations
\begin{eqnarray} \label{fluctuations_Htx_Hty_v1}
&& U \, \Big[ e^{3 \, V + W} H_{tx}'\Big]' -  a_1^2 \, e^{V + W} H_{tx} + 4 \, Q \, U \, a_x' = 0
\nonumber \\[5pt]
&& U \, \Big[ e^{V + 3 \, W} H_{ty}'\Big]' -  a_2^2 \, e^{V + W} H_{ty} + 4 \, Q \, U \, a_y' = 0 
\end{eqnarray}
where we have used that $f' = Q \, e^{-V - W}$. 
Perturbing the equation of motion for the axions in \eqref{Maxwell_axion_EOM} we arrive to the 
following expression
\begin{equation}  \label{axion_perturbation}
\partial_{\mu} \Bigg[\sqrt{-g} \, \left[\frac{1}{2} \, \delta g^{\lambda}{}_{ \lambda}   \, g^{\mu \alpha} \partial_{\alpha} \chi_i + \partial_{\alpha} \delta \chi_i
- \delta g^{\mu \lambda} \, \partial_{\lambda} \chi_i \right] \Bigg] \, = \, 0 \quad {\rm for} \quad i=1,2 \, . 
\end{equation}
Applying \eqref{pert_ansatz} in \eqref{axion_perturbation}, we arrive to two second order differential equations
that are consistent with \eqref{delta_chi_prime}.

The next step in the analysis is coming from the study of the boundary conditions that have to be
imposed on the perturbations at infinity and at the horizon of the black hole. The perturbation of the axion only 
appears in \eqref{delta_chi_prime} and we will assume that $\delta \chi_i$ is analytic at the black hole horizon 
and falls off fast at infinity. To ensure that the perturbation of the gauge field is regular at the black hole horizon, 
we need to change coordinates from $(t,r)$ to Eddington-Finklestein (more details in \cite{Donos:2014cya}).
Regularity at the horizon imply that 
\begin{equation} \label{boundary_condtions_1}
a_x = - \frac{E_x}{4 \, \pi \, T} \, \ln (r-r_h) + {\cal O} (r-r_h) \quad \& \quad 
a_y = - \frac{E_y}{4 \, \pi \, T} \, \ln (r-r_h) + {\cal O} (r-r_h) 
\end{equation}
with the behavior of $U$ close to the horizon being $U \sim 4 \, \pi \, T  (r-r_h)$. Notice that as we approach the 
horizon $a_x' \, U \sim - E_x$ and $a_y' \, U \sim - E_y$.  From \eqref{delta_chi_prime}  
we can see that $H_{xr}$ and $H_{yr}$ are diverging and horizon regularity conditions imply that 
\begin{equation} \label{boundary_condtions_2}
H_{tx} = U \, H_{xr} + {\cal O} (r-r_h) = - \, \frac{4 \, E_x \, Q}{a_1^2\, e^{V +  W}}  + {\cal O} (r-r_h) 
\quad \& \quad 
H_{ty} =  - \, \frac{4 \, E_y \, Q}{a_2^2\, e^{V + W}} + {\cal O} (r-r_h) . 
\end{equation}

The expressions for ${\cal J}^x$ and ${\cal J}^y$ that are defined in \eqref{Jx-Jy_v2} are independent of the 
radial coordinate and we can choose to calculate them at the horizon. Using the boundary conditions 
\eqref{boundary_condtions_1} they become
\begin{equation} \label{Jx-Jy_v2}
 {\cal J}^x = e^{W - V} \, E_x - Q \, H_{t x} \quad \& \quad  
 {\cal J}^y  = e^{V - W} \, E_y - Q\, H_{t y}
\end{equation}
and in order to calculate the conductivity, from the relation ${\cal J} = \sigma \, E$, we have to impose the 
values of the graviton fluctuation $H_{tx}$ and $H_{ty}$ at the horizon. Those are given in \eqref{boundary_condtions_2}  
and they are in agreement with the fluctuation equations \eqref{fluctuations_Htx_Hty_v1}.  Notice that the key point to 
confirm the aforementioned agreement is that the contribution from the first term in 
\eqref{fluctuations_Htx_Hty_v1} close to the horizon vanishes.
Combining \eqref{Jx-Jy_v2} and \eqref{boundary_condtions_2} we can calculate the expressions of the 
conductivities
\begin{equation} \label{xx_yy-conductivities}
\sigma_{xx} = e^{W - V} + \frac{4 \, Q^2}{a_1^2 \, e^{V + W} } \quad \& \quad 
\sigma_{yy} = e^{V - W} + \frac{4 \, Q^2}{a_2^2 \, e^{V + W} } \, . 
\end{equation}
Notice that in the anisotropic case we are considering, i.e. $a_1\ne 0$ \& $a_2 = 0$, the 
$\sigma_{xx}$ of the conductivity is finite while the $\sigma_{yy} $ is infinite.



\bibliographystyle{utphys}

\bibliography{AnisotropicAdSCMT_v2}

\end{document}